预研报告

# 在线广告的跨平台测量方法研究报告

作者：黄楠栖 李纯喜 赵永祥 郭宇春

北京交通大学

2018 年 1 月

# 摘要

  在线广告已经成为主流的广告投放方式。2016 年度，中国在线广告市场规模达到 2902.7 亿元，占整个广告市场收入近七成。

  目前在线广告主要通过广告交易平台（ADX）进行大规模广告投放。广告交易平台实现用户与广告之间的匹配。为了提高广告投放精度，广告交易平台通常在线追踪上网用户数据、挖掘其兴趣，并根据用户兴趣投放相应的广告。

  目前市场上已经出现了大量存在竞争关系的 ADX 平台。对于广告主来说，以下问题至关重要：选择哪个 ADX 平台投放广告才能实现最好的广告投放效果。

  因此，鉴别和比较 ADX 平台的广告投放性能是亟需的基本需求。传统的测量方法通过访问一序列训练网页，在 ADX 平台上构建出一个具有某种兴趣特征的虚拟角色，进而观测 ADX 平台对该虚拟角色投放的广告并评估匹配性能。而这种方法只是对单个 ADX 平台进行测量和性能评估；因为没有统一的测量基准，不适用于多个 ADX 性能的比较研究，因而无法为广告主提供决策参考。为此，本文提出并实现了一个同步跨平台测量方法。该方法能够实现不同 ADX 平台性能的对比，帮助广告主选择合适的广告交易平台。该方法中有如下创新点：

  （1） 提出一个多 ADX 平台性能比较的测量基准。用传统方法独立地测量多个平台时，由于不同平台上训练网页不一致，这些虚拟角色不可避免具有兴趣特征偏差。为了给多个 ADX 平台测量提供统一的测量基准，本文提出了利用多个平台监测的交集网页训练虚拟角色。这种方法可避免单平台测量方法存在的测量偏差。

  （2） 提出并行访问广告链接方法。在传统的测量方法中，需要同时访问网页及网页中的广告内容，但是一旦读取广告内容，会造成平台对虚拟角色的认识偏差。为了解决这个问题，传统方法访问训练网页集合后再获取广告内容，但这种滞后可能导致广告内容变更，从而影响测量精度。本文采用如下方法解决该问题：在获取广告链接的同时用另一个专职读取广告的虚拟角色并发访问这些广告链接。这样，既能实时获取广告链接对应的内容，又不会造成对当前被测虚拟角色认识的偏差。

  为了评估本文提出的同步跨平台测量方法的有效性，设计并开发了一套系统对实际的 ADX 平台进行了测量和比较分析。实验结果达到了预期效果，在同一基准下，该系统能够明显区分出不同 ADX 平台对相同虚拟角色的广告投放差异；并发现不同 ADX 平台广告投放会随虚拟角色改变而变化，且这种变化的灵敏程度存在不同。

**关键词**：在线广告；ADX；跨平台测量

# 目录

# 1　引言

## 1.1　研究背景

　　随着互联网的普及与快速发展，它对人们日常生活有着越来越深入的影响，互联网也演变成除报纸、杂志和电视等传统媒体外一个主要的广告媒介。在线广告从最初比较简单的以展示为主的合约广告开始，在短短十几年的发展过程中，已经演变出根据关键词投放的上下文广告、搜索广告和定向广告，并形成以人群为投放目标、以产品为导向的技术型投放模式。

　　在线广告快速发展，目前已经达到相当大的规模。根据艾瑞咨询 2016 年度中国网络广告核心数据显示，中国在线广告市场规模达到 2902.7 亿元，占五大媒体广告收入近七成，预计至 2018 年整体规模有望突破 6000 亿元[1]。美国网络广告市场规模在 2016 年达到 620 亿美元，远超过报纸杂志等纸质媒体。

　　在线广告传统模式中，只有广告主与媒体主两者参与。所有的广告主都是通过购买媒体网站广告位的形式进行投放在线广告的，上网用户可以在各个媒体网站的预设广告位看到相关广告主打出的广告信息，通过点击行为到达广告落地页，实现推广的目的。但是这种投放方式是十分被动的，广告主是对所有的人进行广告推广，广告主中一直盛行这样一句话："我知道自己的广告浪费了一半，却不知道是哪一半"。据美国国家广告商协会评估[2]，2015 年全球有 70 亿美元的广告由于广告欺诈没有被用户看到而浪费。而用户反映出现自己不感兴趣甚至厌恶的广告或者已经购买过的重复商品广告的现象也很多，这极大的影响了用户的上网体验。

　　为了解决这种针对性不足带来的问题，在线广告系统中出现了实时竞价的模式。实时竞价（RTB）是一种利用在线追踪技术在众多媒体网站中，以每个上网用户个体为投放维度，对广告展示行为进行评估以及出价的竞价技术。在这种广告投放模式下，广告主的广告投放需要在每一次广告展示曝光的基础上进行竞价，竞得者可以获得展现本次广告的机会。

　　在整个 RTB 竞价过程中，ADX 平台作为用户数据的挖掘者，是不可获取的重要角色。ADX 平台通过深入挖掘访问网站用户的行为特征,掌握用户的兴趣偏好，从而根据其兴趣投放相应的广告。当用户浏览网页时，在该网页嵌入广告代码的 ADX 平台可以根据各种在线追踪技术（如 cookie 匹配、指纹识别等）以及平台间的信息共享，来识别用户，然后记录用户的上网行为并分析用户的兴趣偏好，最后根据用户的兴趣向其展示最合适的广告。

## 1.2　研究意义

随着在线广告市场的发展，国内外出现了大量存在竞争关系的 ADX 平台。在国外，主要有雅虎的 RightMedia，谷歌的 DoubleClick 等；在国内，百度联盟、阿里 Tanx、新浪 SAX 等 ADX 平台成为主流平台。另外，苏宁、京东、搜狗、优酷、暴风影音、爱奇艺、芒果广告、触控广告和木瓜移动也都推出了自己的 ADX 平台。

在众多 ADX 平台中，如何选择 ADX 平台对在线广告系统的各个参与者都很重要。ADX 平台的性能对在线广告系统的各个参与方都有直接影响。不具有针对性的广告投放不利于整个在线广告系统。当上网用户浏览网页时，各种他不感兴趣的广告铺天盖地而来，会使他产生反感。这不仅会导致广告点击率没有提高，还会影响用户对该网页的好感，造成网页的受欢迎度下降。相反的，文献[3]研究结果表明，针对用户兴趣爱好进行的广告投放可以改善用户的体验，并提高广告点击率。也就是说，对于上网用户来说，ADX 平台行为定向的效果会改善其上网体验感；对媒体主而言，选择合适的广告平台进行合作对其自身的用户访问量和流量变现也有积极作用；对于广告主，用户对广告的点击直接影响他们的投资回报率；对于 ADX 平台本身，优秀的广告投放性能可以吸引更多与其合作的对象，获得更多的收入。

因此，无论对于广告主、媒体主、用户，或者 ADX 自身而言，鉴别和比较 ADX 平台的广告投放性能是各方都亟需的基本需求。而目前存在的研究没有对 ADX 平台广告投放性能进行比较分析的，因而无法给出参考。本文正是利用现有研究在这方面的空白，提出了一套多 ADX 平台测量系统，比较不同 ADX 平台广告投放效果的差异。利用该系统对多个 ADX 平台进行测量，可以为广告主和媒体提供比较直观的广告效果判断，从而更有效地利用在线广告获利。该评估体系的建立不仅能使广告主、广告平台的广告投放效果优化、投资获利最大化，还能改善用户上网体验，优化互联网广告生态圈环境，从而促进以广告收入为主要营收的互联网的健康发展。

## 1.3　国内外研究现状

自从在线广告出现以来，在线广告系统受到工业界的高度追捧，学术界也对其密切关注并做了大量研究。

早在 2010 年，来自微软的 Saikat Guha 和就职于 MPI-SWS 的 Bin Cheng, Paul Francis[45]就发表了一篇研究在线广告系统的文章。他们设计出一套系统地测量 Google 在线广告的方法，并且对用户行为是否影响搜索页面上的广告、历史搜索

是否影响浏览页面上的广告，以及用户的哪些数据会影响社交网络上的广告做出分析。这是现有研究中最早对 ADX 平台进行研究的，但其结果显示用户行为对搜索广告没有影响；谷歌没有使用近期浏览、搜索或点击行为来定向广告。

Barford 等人对在线广告系统有更广泛的研究，他们观察谁是最主要的 ADX 平台、哪部分广告是定向广告，以及用户的哪些特征会导致定向[51]。这些结论为后来的在线行为广告和广告系统研究提供了很大参考。在接下来不到十年的时间里，在线广告系统领域有了很多发现。

近年来，研究者对在线广告系统的研究主要集中在行为定向广告上。Mikians 和 Cuevas 等提出了一个测试在线行为广告出现频率和展示效果的方法，并将该方法应用于可扩展测量系统中，进行大规模实验。最终他们发现行为定向广告是在线广告中常用的技术，而且用户收到的行为定向广告数量与其行为或者兴趣标签的经济价值成正相关。

类似地，为了研究用户行为对在线广告的影响，Yan Jun 和 Liu Ning[53]等通过收集商业搜索引擎上的日志数据，分析了 7 天时间里约 642 万用户对 33 万多广告的点击。通过比较对用户分类、基于用户浏览行为定向和基于用户搜索行为定向这几种不同策略下的广告点击率，他们得出了几个结论：点击相同广告的用户相似度高；对用户分类并进行行为定向可将广告点击率提高六倍多；以及基于用户短期内的搜索行为定向广告效果更好，从而验证了行为定向对广告效果的有利影响。Bahtiar[54]等则延伸地研究在线行为定向广告对用户购买意向的影响，这也是对行为定向广告的转化效果的扩展研究。

在现有文献里，除了文献[45]针对谷歌 ADX 平台进行了测量，其他文献都没有对 ADX 平台进行研究。但因该研究是在定向广告还未流行的 2010 年所做，其得出的结论是谷歌的在线广告不受用户行为影响。此外，并没有研究对多个 ADX 平台进行测量。由此可见，现有研究缺乏对多个 ADX 平台的鉴别研究。

## 1.4  研究内容

为了鉴别和比较多个 ADX 平台广告投放性能，本文设计了一套同步跨平台测量方法。其基本思想是：通过访问多个 ADX 平台共同监测的网页，在这些平台上构建出一个具有某种兴趣特征的虚拟角色，并获取不同 ADX 平台针对该角色投放的广告；对这些广告进行比较分析，从而评估和比较不同 ADX 平台的广告投放性能。该方法中有如下创新点：

（1）利用多个平台监测的交集网页训练虚拟角色。为了比较多个平台，传统的测量方法在各个平台单独塑造兴趣特征相同的虚拟角色，由于不同平台上训练

网页不一致,这些虚拟角色不可避免具有不同特征偏差。为了给多个 ADX 平台测量提供统一的测量基准,本文利用多个平台监测网页的交集训练虚拟角色可避免这种偏差。

(2) 提出并行访问广告链接方法。在获取广告链接的同时用另一个专职读取广告的虚拟角色并发访问这些广告链接。这样,既能实时获取广告链接对应的内容,又不会造成对当前被测虚拟角色认识的偏差。

为了评估本文提出的同步跨平台测量方法的有效性,设计并开发了一套系统对实际的 ADX 平台进行了测量和比较分析。实验结果达到了预期效果,在同一基准下,该系统能够明显区分出不同 ADX 平台对相同虚拟角色的广告投放差异;并发现不同 ADX 平台广告投放会随虚拟角色改变而变化,且这种变化的灵敏程度存在不同。

## 1.5 文章组织结构

本文一共分为五章,每章的主要内容包括:

第一章为"引言",主要介绍了本文的研究背景和研究意义,最后概括了本文的主要研究工作及内容安排。

第二章介绍了在线广告系统的相关研究,主要包括在线广告生态系统的系统组成、所用的相关技术以及现有的研究内容,并总结了现有研究的优缺点。

第三章介绍了本文提出的同步跨平台测量方法,论述了如何实现该方法,包括设计思想和设计流程。

第四章利用本文提出的方法对多个 ADX 平台进行性能评估,比较分析了谷歌及国内几大广告交易平台广告投放的效果。

第五章为结论部分,总结了本文的研究内容,并提出未来研究工作的展望。

# 2　相关研究

为了更清晰地了解本文研究的应用场景，本章第一小节介绍在线广告生态系统，第二小节介绍行为定向广告相关技术，第三小节对现有研究进行总结和对比，并阐述与本文的联系。

## 2.1　在线广告生态系统

本小结介绍了在线广告生态系统，为读者能够理解本文主要技术提供理论基础。第一部分从经济和技术两方面介绍了在线广告生态系统的框架；第二部分对在线广告系统中的主要角色做了整体介绍，第三部分描述了在线广告服务的原理，最后两部分详细介绍了定向广告以及本文所研究的行为定向广告。

### 2.1.1　系统架构

在线广告，也称互联网广告，顾名思义，指的是在互联网媒体上投放的广告。因为消费者/用户在网上花费的时间日渐增长，在线广告对于品牌商和广告主的吸引力也越来越大。在线广告（特别是消费品）的一个优势在于它可以将广告活动及其相关成本与产品销售相关联，例如通过跟踪对在线广告的点击可以推断出该品牌的网上店铺的购买活动

整个在线广告系统的结构如图 2-1 所示，接下来将详细介绍该系统中的主要参与者以及他们扮演的角色。

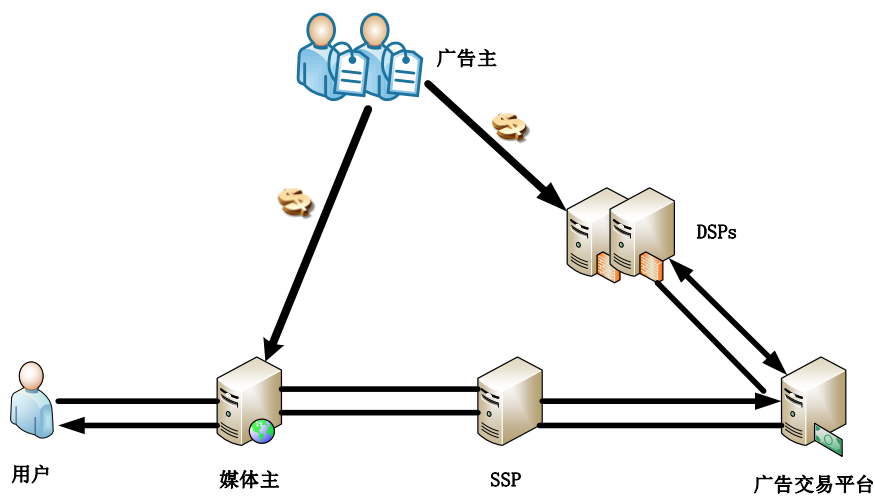

图 2-1 在线广告系统结构图

Figure 2-1 System architecture of online advertising

在线广告系统是由众多具有特殊以及互补作用的实体共同组成的，他们最终的目的是在互联网上展示广告。媒体主、广告主、广告平台、需求方平台和供给方平台都是广告投放过程的参与者。尽管在线广告系统相当复杂且在不断演变，但是目前在线广告的投放主要是由媒体主、广告主和广告交易平台三方主导的，接下来将详细介绍在线广告系统的各个参与者：

（1） 媒体主

媒体主是一些拥有网页（或网站）的实体，他们已经取得较大的用户访问量（在线广告系统中称为流量），为了将这些线上流量转变成现金收益，他们在网站的 HTML 页面上插入广告位。例如，纽约时报、AOL 等就是典型的媒体主。

（2） 广告主

广告主是希望通过向媒体主（传统模式）或广告交易平台（RTB 模式）付费，在媒体主提供的广告位上展示自己广告的实体。为了实现此目标，广告主通常与一个或多个 广告交易平台合作。正如本文将在 2.1.2 节中介绍的，目前有两种广告交易模式，而在这两种模式中广告主扮演的角色也不一样。在传统模式中，广告主直接与其产品相关度高或者流量大的媒体主协商并投放广告。而在目前比较流行的广告网络模式中，广告主通过与广告交易平台合作而不直接与媒体主打交道，他们只需要向广告交易平台指明适合其广告产品的定向目标，即他们希望对其展示广告的用户。例如，广告主可能希望广告平台向那些对足球感兴趣的人或居住在北京的人们提供广告。广告主还必须指定他们愿意为每次广告显示和用户点击支付的费用。

（3） 广告交易平台

广告交易平台（Ad Exchange，以下简称 ADX）是将媒体主与广告主联系起来的实体，即它将广告主的广告投放到媒体主提供的广告位上。为了将广告主的广告精准投放到目标人群，ADX 平台通常会追踪并刻画用户画像[4]（包括用户背景、特征、性格标签、行为场景等），从而针对用户的兴趣、地理位置或其他个人数据进行广告投放。 ADX 平台分别通过 SSP 和 DSP 为媒体主和广告主服务。SSP 是供给方平台，是将大量广告位提供者（也就是媒体主）聚集到一起集中管理的平台。DSP 是需求方平台，其主要任务是按照定制化的人群标签购买广告位。

（4） 品牌商与用户

在线广告系统中，除了上面三个主要角色外，还有两个重要群体：品牌商和用户。品牌商向广告主付费以帮助他们销售自己的产品和服务。用户是在线广告系统的被动参与方，是所有在线广告的最终目标。

## 2.1.2 广告投放机制

上一小节已经介绍了在线广告系统的主要参与者及其扮演的角色，接下来将介绍在线广告系统两种广告投放模式：传统模式和实时竞价（RTB）模式，并分别介绍每种模式下的广告投放机制。

（1） 传统模式

在线广告系统传统模式是只有广告主与媒体主两者参与的模式。广告主和媒体主直接进行协商，没有 ADX 平台的参与。在这种模式中，流量较大的网站直接面向广告主销售广告位。通常来说，这些广告是非定向的，广告产品往往与该网站的内容相关。这主要是因为媒体主跟踪和刻画用户的范围仅限于自己的网站，也许还有几个合作伙伴。例如，纽约时报 Web 站点也可以在国际先驱报、论坛报等和它同属一个媒体集团的网站上追踪用户，但是这样的跟踪能力，与可在 200 万网站上追踪用户的谷歌广告平台比起来是微乎其微的。

在线广告市场上有多种广告计费模式，广告主可与媒体主协商采用哪种支付方式向其付费。目前有几种常用的计费方式：CPM 结算，即按照千次展示结算，是广告主与媒体主约定好千次展示的计费标准；CPC 结算，即按点击结算，广告主在用户产生点击行为时付费；CPS 结算，即按照销售订单数来结算，广告主只按照最后的收益来结算；CPT 结算，一般是针对大品牌广告主按独占时间段收取费用的方式[5]。广告主的目标是在他们认为吸引品牌目标人群的网站上投放广告。因此，他们面临的挑战是如何确定这些网站并促进广告投放。

（2） 实时竞价模式

除了直接与媒体主协商之外，广告主经常与 ADX 平台合作，通过 ADX 平台投放广告。ADX 平台采用实时竞价方式投放广告，因此这种有 ADX 平台参与的模式称为实时竞价模式。

ADX 平台的参与对广告主和媒体主都起到有利作用。随着媒体网站数量越来越大且日益多样化，广告主很难从中挑选出他们认为吸引品牌目标人群的网站。与 ADX 平台合作后，广告主就可以利用 ADX 平台提供的用户画像自主选择媒体流量。另一方面，对于媒体主来说，只靠他们与广告主直接接洽可能无法售卖出自己的所有广告位，而 ADX 平台的加入则解决了这个问题。

由于 ADX 平台对在线广告系统有积极作用，自从 2005 年首个 ADX 平台 RightMedia 出现[6]，通过 ADX 平台进行广告交易逐渐成为最流行的方式。ADX 平台的业务范围相当广泛，其中最主要的就是定向广告。据文献[7,8]显示，Google Ad Exchange 是在线广告领域使用最广泛的广告平台之一。国内最为人熟知的 ADX 平台有百度的 BES、阿里妈妈的 TANX、新浪（SAX）。另外，苏宁、京东、搜狗、优酷、暴风影音、爱奇艺、芒果广告、触控广告和木瓜移动也都推出了自己的 ADX 平台。

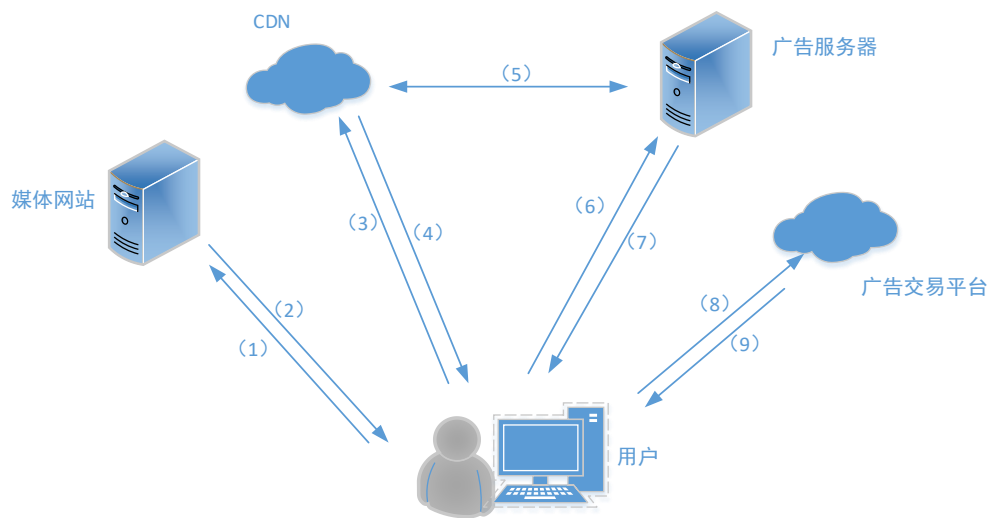

图 2-2 RTB 模式数据交换图

Figure 2-2 RTB mode data exchange diagram

在实时竞价模式中，在线广告生态系统的参与者之间有大量潜在的数据交换。图 2-2 描述了一个简单的例子。整个广告投送过程始于媒体主在其网站上嵌入合作 ADX 平台的链接。如图 2-2 所示，当用户访问一个网页时，用户浏览器向该网页服务器发送加载网页的请求（1）。服务器除了向浏览器返回基本页面，还会告诉浏览器它需要加载广告代码（2）。浏览器收到通知后，向内容分发网络（CDN）发送加载广告代码请求（3）。CDN 是构建在网络之上的内容分发网络，为了使用户可就近取得所需内容、提高用户访问网站的响应速度,广告代码通常存储在CDN 中。广告代码是简单的 HREF 字符串，它是一种指定超链接目标的 URL，通常是引用嵌入在 CDN 基础架构中的 JavaScript 代码。CDN 将广告代码发送给用户浏览器（4），并且会和广告服务器联系即时更新自己网络中广告代码（5）。接下来，用户浏览器向广告服务器发送请求加载广告内容（6），广告服务器将其重定向至 ADX 平台（7）（8），最终 ADX 平台向浏览器返回广告并在用户浏览页面上显示。为了确保良好的用户体验，整个过程必须快速进行（通常大约为几十毫秒）。

在整个广告投放过程中， ADX 平台可通过使用第三方 cookie、网络指纹或其他跟踪技术（详见 2.2 节），跟踪访问这个网站和与它合作的任何其他站点的用户。这种跟踪用户的能力对广告平台至关重要。它使广告平台能够了解用户正在访问的网页及其内容、IP 地址以及最重要的用户的 Web 浏览兴趣。

所有这些关于用户的重要数据都被广告平台用来为目标广告服务。为了完成广告服务，绝大多数的广告平台都开发出自己的一套广告投放算法。上述用户数据和所有广告主显示广告的目标和预算都是这些算法的输入，这些算法负责选择在特定广告位显示何种广告。显然，他们的主要目标是最大限度地提高广告平台的收

入，同时满足广告主的需求。

### 2.1.3 定向广告

定向广告中的"定向"实际上是对受众的筛选，即通过技术手段标定某个用户的性别、年龄或其他标签。定向广告也就是指根据用户的这些标签相应地投放广告。

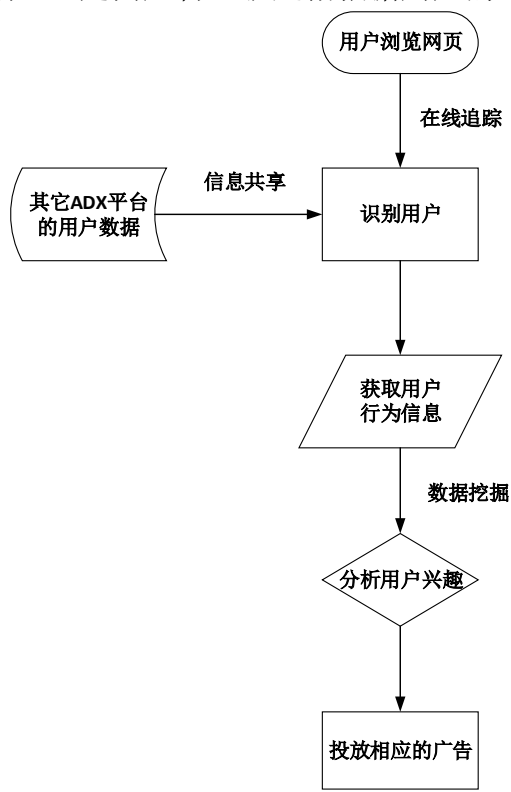

图 2-3 行为定向广告流程图

Figure 2-3 The workflow diagram of behavioral targeting advertising

根据广告投放者选择的定向目标不同，市场上出现了多种定向方式。目前最受欢迎的定向目标包括用户正在访问的网页、用户位置和用户浏览兴趣。接下来，将介绍几种比较流行的定向方式[5]：

（1） 上下文定向

广告系统通过分析语义和语境，向用户展示与其访问网页内容相关的广告，这就是上下文定向。上下文定向的粒度可以是关键词、主题，也可以是根据广告主需求确定的分类。这种广告投放策略的一个典型例子就是，广告系统将一家医疗保险公司的广告投放在"健康与健身"类别的网站上。

（2） 地理位置定向

这是一种很直觉的定向方式，它们通常是根据用户的位置（例如，由他们智能手机或平板电脑的 GPS，以及用户设备 Wi-Fi 接入点和 IP 地址）来投放广告。举

个简单的例子，假设某电商网站只在北京运营和送货，那么其在线广告一般应该定向在北京的用户。

（3） 行为定向

广告系统还可以根据用户的各种上网行为来向用户投放广告。通常情况下，通过追踪并收集用户的上网信息，广告平台和第三方可以推断出用户的兴趣偏好，从而根据其兴趣偏好投放广告。

（4） 重定向

这是一种最简单的定制化标签，其原理是对某个广告主过去一段时间的访客投放广告以提升效果。重定向被公认为是精准度最高、效果最突出的定向方式，不过其人群覆盖量往往较小。

最后，通过给出广告主如何设置定向广告的真实示例来总结本小节。图 2-4 显示了百度广告交易平台上的定向设置面板，广告主可以根据年龄、性别、学历等基本信息以及用户环境等定义目标受众。由于广告预算的限制，对于每个广告系列，广告主必须适当地配置所有这些变量。

图 2-4 百度广告交易平台定向设置面板

Figure 2-4 Baidu ADX orientation settings panel

## 2.1.4 行为定向广告

行为定向是在线广告中非常重要的一种定向方式，其框架是根据用户的历史访问行为了解用户兴趣，从而投送相关广告。用户的搜索（Search）、分享（Share）、页面浏览（Page view）及广告浏览（Ad view）等在线行为可以被广告系统广泛采集，并且对于受众定向或广告决策有明确作用。基于用户的上网行为，ADX 平台可以构建用户画像，并且可以使用这些用户画像来提高广告的有效性[4]，从而导致更高的点击率转化率[5]，从而给媒体主、广告主带来更多的收入。

在用户的多种在线行为中，网页浏览是用户在目的比较弱的网上冲浪过程中产生的，其所涉及的兴趣领域对把握用户信息有价值。虽然精准度有限，但是由于其数据量是各种用户行为中最大的，往往被用作推测用户偏好的参考信息。基于此，为了测量行为定向广告，并评估不同 ADX 平台行为定向的效果，本文以用户的页面浏览行为为依据构建基于偏好的虚拟人。本文所述构建虚拟人并非创造出真实存在的实体，而是利用一系列网页浏览行为使 ADX 平台认为存在一个有某种偏好的用户。

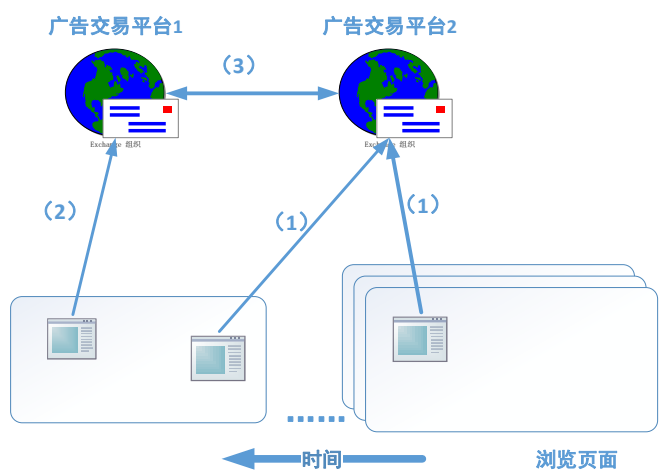

图 2-5 行为定向广告示意图

Figure 2-5 Behavioral targeted advertising schematic

图 2-5 概述了 ADX 平台参与下行为定向广告投放流程，以及该过程中用户数据如何被利用的。用户张三在上网过程中，会访问一系列网页，这时他正在被这些页面上存在的多个 ADX 平台跟踪（1）。当张三浏览另一新网页时，该网页上的某个 ADX 平台可能已经通过之前的网页跟踪了他（1），并根据他们掌握的关于他的信息，相应地刻画他。另一种可能的情况是 ADX 平台出现在张三当前访问的页面上，但并没有出现在她之前访问的网页上（2）。在这种情况下，该 ADX 平台可以显示随机广告（非定向）或从其他 ADX 平台那获取关于张三的信息以显示定向广告（3）。

## 2.2　在线追踪技术

本小节主要介绍了在线追踪技术，目的是为读者理解 ADX 平台如何追踪用户、本文如何研究的行为定向广告场景打下技术基础，理解怎么追踪，怎么构建虚拟人，怎么评估平台性能。

在线追踪是指用户在上网时产生的行为被记录。在线追踪涉及三个参与者：上网用户、用户访问的网站以及记录用户浏览行为的第三方。

根据第三方的目的不同，第三方可分为两类。一类第三方的目的是收集用户数据，例如 ADX 平台和社交网络插件，他们利用用户数据投放针对性的广告（ADX 平台）或者其他形式的个性化（社交媒体），从而获得高额收入[9,10]。另一方面，内容分发网络作为第三方，其主要目的是加速网站加载，属于另一类第三方[18,19]。

为了获取用户数据， ADX 平台这类第三方使用多种追踪技术对用户进行追踪。接下来介绍几种常见的第三方追踪技术：基于 HTTP 请求、Cookie 匹配和指纹识别。

### 2.2.1　基于 HTTP 请求

当用户访问一个网站时，他所使用的浏览器向该网站的服务器发出一个 HTTP \ GET 请求，服务器返回一个包含该网页 HTML 代码的 HTTP 响应，通常返回的 HTML 代码中包含对第三方的引用，例如样式表，JavaScript 代码和在浏览器中呈现页面所需的图像。接下来，用户的浏览器会自动发出对这些第三方的请求。

根据用户浏览器对第三方服务器的请求，第三方可获取并记录关于用户的各种信息。这些信息包括发出请求的计算机的 IP 地址、请求的日期和时间、用户所使用的计算机和 Web 浏览器的类型（称为用户代理字段）以及发起请求的页面的地址，称为源地址（REFERER）[11]。图 2-6 显示了一个利用 HTTP 请求获取用户数据的例子。由图中信息可知，第三方服务器接收到名为 tracking_pixel.png 的文件请求，根据请求信息它可以知道一个 IP 地址为 8.67.53.09 的用户正在访问 http://example.com/private_matters.html 网页，并且该用户是在 Macintosh 计算机上使用 Safari 浏览器浏览网页。

```
IP: 8.67.53.09
DATE: [10/May/2014:19:54:25 +0000]
REQUEST: "GET /tracking_pixel.png HTTP/1.1"
REFERER: "http://example.com/private_matters.html"
USER-AGENT: "Macintosh; Intel Mac OS X 10_8_5...AppleWebKit/537.71"
```

图 2-6 HTTP 请求数据图

Figure 2-6 HTTP request data

当第三方服务器收到很多这样的记录后，它可以为属于相同 IP 地址和 USER-AGENT 的信息组合建立行为模式。这就是第三方追踪最基本的形式，并且适用于 web 上的所有 HTTP 请求。

为了获取用户的 HTTP 请求数据，第三方必须先将自己的追踪元素嵌入在网页的源代码中。在大多数情况下，第三方追踪者向媒体主支付一定的费用，从而在媒体主网页中嵌入追踪代码，这也正是媒体主获利的主要业务模式[12]。

### 2.2.2　Cookie 匹配

除了利用 HTTP 请求数据追踪用户外，Cookie 匹配也被用于第三方追踪。某些第三方追踪者使用称为 cookie 的小文件在用户的计算机上存储信息[13]，cookie 可以被用作固定在用户浏览器上的唯一标识符[14]。它的作用类似于鸟类学家安放在候鸟上的追踪手镯，即使用户访问不相关的网站，跟踪 cookie 也会将其标记并追踪。例如，当用户访问 www.cnn.com 时，浏览器可以向 doubleclick.net 发出额外的请求，以加载有针对性的广告，并且向 facebook.com 发送请求加载"赞"按钮，于是 Doubleclick 和 Facebook 都了解到该用户对 CNN 的访问。

### 2.2.3　指纹识别

虽然基于 cookie 的追踪非常普遍[15]，其他类型的追踪技术也相继出现。包括使用其他客户端存储机制，如 HTML5 LocalStorage，或使用浏览器和/或机器指纹识别来重新识别用户而不需要存储本地状态[16,17]。指纹识别是一种高级的追踪形式[18]，通过检测计算机独特特征来追踪用户。举个例子，安装在计算机上的特定字体表就像手指上的圆环和螺纹的组合一样通常是独一无二的，它就像指纹一样，想要改变或隐藏起来都是很困难的。相比于 cookie 可能会被用户删除或阻止，指纹识别追踪能力更稳定也更厉害。正因如此，越来越多第三方的注意力正在迅速转向指纹识别技术。

## 2.3　现有研究总结

互联网广告的飞速发展引起了人们对其效果的兴趣。互联网广告是否有效、对用户行为有多大影响，这些问题已经吸引一大批国内外工业界和学术界的研究人员的关注和讨论。

由于互联网广告具有互动性和个性化等传统媒介所不具有的特性，目前网络广告还没有建立起一套比较直观可靠的效果评估体系。但是工业界和学术界的研究人员对在线广告效果的探究从未停止而且涉及到广泛内容。

现有研究主要从三个方面对在线广告系统进行研究：追踪技术、用户画像和广告系统测量。本节将分别从这几个方面总结现有技术和测量方法的优缺点，以及和本文工作的联系。

### 2.3.1 追踪与追踪机制

（1）相关文献

为了提高广告定向能力，广告投放者必须广泛地跟踪用户。对于广告投放者的追踪机制，一部分研究者对其进行了研究。Krishnamurthy 等人统计了第三方追踪的普遍度，并对第三方追踪给用户带来的隐私影响进行了评估[19,20,21]。Roesner 等人开发了一套分类体系，该体系研究了多种跟踪机制（例如，HTML5 LocalStorage 和 Flash LSO）以及阻止这些机制的策略[22]。

Cookie 匹配作为一种重要的追踪技术，一直受到研究者的广泛关注。Acar 等人发现，在 Alexa Top-3K 网站中，有数百个网站通过 cookie 匹配向彼此传递了用户标识符[23]。Falahrastegar 等人发现有一类网站利用唯一的、匹配的 cookie 共享用户数据[24]。对于各网站之间是否愿意进行 cookie 匹配，以及 cookie 匹配能否帮助网站获取更大利益，Ghosh 等用博弈论进行了分析。他们发现，当不同广告主拥有同一类目标人群时，帮他们展示广告的媒体网站对 cookie 匹配的态度是一致的。也就是说各网站要么都愿意进行 cookie 匹配，要么都不同意。当不同广告主目标人群相差较大时，各网站之间进行 cookie 匹配会导致信息泄露。例如，网站 1 上有 H 类用户，网站 2 上有 L 类用户，广告主的目标人群是 H 类用户。那么网站 1 与网站 2 进行 cookie 匹配，可帮助网站 2 获取 H 类用户数据，在自己网站上区分 H 类和 L 类用户并定向投放广告。这样可以吸引广告主在网站 2 上投放广告，而不完全依赖于网站 1。因此，通过 cookie 匹配，网站 2 的收入提高了，网站 1 的收入却减少了[25]。除了媒体网站之间存在 cookie 匹配，Olejnik 等人发现了 125 个 ADX 平台也使用 Cookie 匹配。此外，他们还通过参与广告竞标研究了 RTB 广告的投放机制[26,27]。

多项研究发现，追踪机制的不断发展使得用户难以防备。尽管用户可以通过清除 Cookie 或使用隐身浏览模式来避开追踪者，但不少第三方已经使用 Evercookies 和指纹识别等技术进行了反击。 Evercookies 在浏览器中的许多地方存储了跟踪者的状态（例如，FlashLSOs，etags 等），因此即使用户删除了他们的 cookie，他们也

能利用存储的信息重新生成跟踪标识符[28,29,30,31]。指纹识别包括根据用户的浏览器[32,33]，浏览历史[34]和计算机（如 HTML5 画布[35]）的特性为用户生成唯一的 ID。一些研究已经发现了使用指纹识别技术的追踪者[36,38,39]；Nikiforakis 等人建议可以通过有意向用户的浏览器添加更多的信息噪声来阻止指纹识别[37]。

不过，也有研究给出了相反的观点。前一段时间，Cahn 等人对整个网络中的 cookie 特征进行了广泛的调查，发现只有不到 1% 的追踪者能够在 Alexa Top-10K[40] 中的 75% 的网站上收集信息[41]。Li 等声称大多数跟踪 cookie 可以使用简单的机器学习方法自动检测[42]。

追踪技术的使用不仅对广告投放者和媒体网站有积极效应，还能给追踪者本身带来巨大的收益。Gill 等人收集了数百万上网用户的 HTTP 记录，研究各个 ADX 平台收集了多少用户信息，以及这些信息对定向广告的价值有多大。他们开发了一个简单模型，来捕捉在线广告收入的各种参数，这些参数的值是通过 HTTP 跟踪来估计的。他们的研究结果显示，不同 ADX 平台的收入存在较大差异，5% 的 ADX 平台收入占整个行业收入的 90%。Google 无论是在收入还是覆盖率方面都占据主导地位（它出现在 80% 的媒体网站上）。不过他们也表明，如果对广告收入贡献最大的前 5% 的用户启动隐私保护，并且媒体主没有做出相应的反应，那么 ADX 平台的收入会减少 30%[43]。

（2）小结

现有研究发现，目前存在的追踪技术主要包括基于 HTTP 请求、Cookie 匹配和指纹识别技术。基于 HTTP 请求是目前在线广告系统中应用范围最广的追踪技术，ADX 平台通过将广告代码嵌入媒体主的网页上即可对访问该页面的用户进行追踪。Cookie 匹配技术不仅可以利用 cookie 存储用户信息，还能在多个 ADX 平台之间实现信息交换和共享。相比于 cookie 可能会被用户删除或阻止，指纹识别技术更具有独特性且难以被隐藏，是目前追踪能力最强的追踪技术。本文根据基于 HTTP 请求追踪技术的原理，对网站上的 ADX 平台进行识别和筛选。因为其应用范围最广，符合本文大规模测量的研究方法。

## 2.3.2　用户画像

（1）相关文献

通过使用多种追踪技术获取用户数据，广告系统可根据这些数据对用户画像。为了研究 ADX 平台如何构建用户画像，有几个文献以流行程度较高的谷歌平台的追踪数据作为研究对象[44]。谷歌平台的广告偏好管理器（Ad Preferences Manager）是为保护用户隐私设计的一个工具，用户可以从中看到谷歌对其基本信息和兴趣

偏好的了解程度，并自行定义感兴趣并希望与广告相关联的关键词。利用该工具，一些研究者以用户基本信息（地理位置和年龄等）和上网行为（浏览页面、搜索关键词等）作为输入，谷歌广告偏好管理器的信息和展示广告作为输出，研究谷歌如何利用获取的用户数据刻画用户的[45,46]。然而，另外两项研究发现，谷歌广告偏好器并没有把自己知道的所有信息告知用户[47,48]。这一研究发现，无疑给以之前使用谷歌广告偏好管理器数据的研究结果产生不利影响。

为了对抗这种透明度的不足，Lecuyer 等人构建了一个基于可扩展实验和统计分析的系统。通过测量分析输入哪些信息会导致哪些输出，研究 Gmail 和互联网中的定向广告，并发现谷歌声称没有针对敏感和被禁止的话题这一点不符合事实[49,50]。

（2）小结

现有研究中用户画像最常采用的方法是创建模拟用户上网行为（浏览页面、搜索关键词等）的客户端。对于 ADX 平台来说，用户画像是用于刻画用户偏好从而定向投放广告的。为了探测 ADX 平台是如何构建用户画像、怎么利用用户画像定向投放广告以及定向广告的投放效果，研究者需要创造大量具有行为偏好的虚拟用户，让 ADX 平台对其定向投放广告。这种方法是一种比较准确且可扩展实验的，本文就是在此基础上创建具有行为偏好的虚拟人，从而探测 ADX 平台行为定向效果。

### 2.3.3 广告系统测量

（1）相关文献

自从在线广告出现以来，在线广告系统受到工业界的高度追捧，学术界也对其密切关注并做了大量研究。早在 2010 年，来自微软的 Saikat Guha 和就职于 MPI-SWS 的 Bin Cheng, Paul Francis[45]就发表了一篇研究在线广告系统的文章。他们设计出一套系统地测量 Google 在线广告的方法，并且对用户行为是否影响搜索页面上的广告、历史搜索是否影响浏览页面上的广告，以及用户的哪些数据会影响社交网络上的广告做出分析。其分析结果显示用户行为对搜索广告没有影响；谷歌没有使用近期浏览、搜索或点击行为来定向广告；用户的爱好、教育程度等个人信息对 Facebook 上的广告投放都有影响，但是性别和年龄是最大影响因素。他们指出测量定向广告的困难在于广告流失（旧广告的消失和新广告的出现）产生的噪声干扰、网络系统带来的影响（如 DNS 负载均衡导致客户端访问的服务器不同，从而收到的广告不同）以及时间变化带来的影响（显示广告随时间变化）。

Barford 等人对在线广告系统有更广泛的研究，他们观察谁是最主要的 ADX

平台、哪部分广告是定向广告，以及用户的哪些特征会导致定向[51]。这些结论为后来的在线行为广告和广告系统研究提供了很大参考。在接下来不到十年的时间里，在线广告系统领域有了很多发现。

近年来，研究者对在线广告系统的研究主要集中在行为定向广告上。Mikians 和 Cuevas 等提出了一个测试在线行为广告出现频率和展示效果的方法，并将该方法应用于可扩展测量系统中，进行大规模实验。最终他们发现行为定向广告是在线广告中常用的技术，而且用户收到的行为定向广告数量与其行为或者兴趣标签的经济价值成正相关。也就是说，如果用户的行为偏好或兴趣标签被认为是比较昂贵的，那么他们会比其他人收到更多行为定向广告。更重要的一点是，他们还发现行为定向广告和健康、政治等这类敏感话题也有关联[52]。

类似地，为了研究用户行为对在线广告的影响，Yan Jun 和 Liu Ning[53]等通过收集商业搜索引擎上的日志数据，分析了 7 天时间里约 642 万用户对 33 万多广告的点击。通过比较对用户分类、基于用户浏览行为定向和基于用户搜索行为定向这几种不同策略下的广告点击率，他们得出了几个结论：点击相同广告的用户相似度高；对用户分类并进行行为定向可将广告点击率提高六倍多；以及基于用户短期内的搜索行为定向广告效果更好，从而验证了行为定向对广告效果的有利影响。Bahtiar[54]等则延伸地研究在线行为定向广告对用户购买意向的影响，这也是对行为定向广告的转化效果的扩展研究。

此外，实时竞价广告作为在线广告系统中的重要组成，一部分人对其竞价策略进行了研究。Yuan Shuai 和 Wang Jun 等人利用某 ADX 平台提供的需求方和供给方的数据，观测实时竞价广告模式中广告位的售卖方式和交易中的竞价机制。从研究中，他们发现各种统计数据都具有周期性，包括广告展示次数、点击次数、出价和转换率（观看后和点击后），这表明时间依赖模型适合用于捕获 RTB 中的重复数据。他们还发现，尽管 ADX 平台都声称竞价广告是按广义第二高价成交，实际上按最高价成交的广告收入占总收入的 55.4%。因此，他们认为，现行实时竞价系统中 soft floor price 的设定是对广告主不利的，因为当最高出价小于 soft floor price 时广告交易以最高价成交。此外，他们对交易率的分析表明，当前的竞价策略远不是最优的。这也表明研究出一种结合用户实时行为、广告的展示频率和展示时间长短的优化算法来优化竞价策略的必要性[55]。

在线广告系统的快速发展源于其带来的良好经济效益。Gill 等人开发了一个简单模型，来捕捉在线广告收入的各种参数，这些参数的值是通过 HTTP 跟踪来估计的。他们的研究结果显示，不同 ADX 平台的收入存在较大差异，5%的 ADX 平台收入占整个行业收入的 90%。其中，Google 无论是在收入还是覆盖率方面都占据主导地位（它出现在 80%的媒体网站上）。不过他们也表明，如果对广告收入

贡献最大的前 5％的用户启动隐私保护，并且媒体主没有做出相应的反应，那么 ADX 平台的收入会减少 30％[43]。然而，另一方面，Laussel 等分析了从非定向广告到定向广告转变过程中投放广告的公司是否能获得更多利润。他们指出投放广告的公司在广告价格不同时考虑的市场角度不同，而采用定向广告并不一定能为公司带来更大利润，为研究定向广告效果提供了辩证的观点[56]。

另外，对于在线广告带来的隐私问题，也有研究者做过测量。一篇较早期的文献里，Korolova 等人发现了一种利用广告系统提供的定向功能破坏用户隐私的新型攻击。他们研究了全球最大的在线社交网络 Facebook 提供的广告系统，以及系统设计给用户隐私带来的风险。他们提出几个新的方法，并找到了攻击者利用广告系统获得用户私人信息的实际证据[57]。

由于移动互联网越来越显著的重要性，移动广告也越来越被关注。Rodriguez 等人研究了移动设备上的广告生态系统。在这项研究中，他们利用拥有 300 多万用户的欧洲主要移动运营商提供的匿名数据集（包含一天的流量），对移动广告进行多个维度测量，例如移动广告总流量、展示频率以及对移动设备电量潜在的影响等。他们发现目前移动广告投放中存在一些效率极低的例子。并且讨论了利用一些常见的技术（如预取和缓存），可以减少目前系统中的在能源和网络信号上的开销。通过 Android 设备原型实现，他们发现即使包含离线广告的 App 限制了广告相关的流量，手机的能耗仍然提高了 50％[58]。

（2）小结

现有研究对在线广告系统的多个方面进行了探索，但仍有一定局限性。

不少已有文献研究了在线广告投放是否参照了用户上网行为。文献[52,53,54,58]研究了在线广告系统中是否存在行为定向广告，以及行为定向对在线广告的点击率、转化率的影响。

对于行为定向广告对在线广告系统经济效益的影响，现有研究得出了不同结论。文献[43]发现在线广告给 ADX 平台带来了巨大收入。相反地，文献[56]却认为定向广告并不能给广告主带来更多利润。

值得注意的是，以上文献都没有对 ADX 平台进行研究。在现有文献里，只有文献[45]针对谷歌 ADX 平台进行了测量，但因该研究是在定向广告还未流行的 2010 年所做，其得出的结论是谷歌的在线广告不受用户行为影响。此外，在测量方法上，本文针对文献[45]中测量方法存在的局限性进行了改进。文献[45]中针对单个平台进行测量，所使用的方法为：先定义某个虚拟角色，在平台监测的网页中找到与虚拟角色特匹配的网页；通过多次访问这些网页，使该平台认识到有这个特定的虚拟角色存在；进一步，收集该平台针对该虚拟角色投放的广告，测量这些广告与该虚拟角色兴趣特征的相关性。而本文所采用的方法适用于对多个平台的同步

测量，主要体现在：利用多个平台监测的交集网页训练虚拟角色；提出一种并行访问广告链接方法以及使用基于图像识别的广告内容提取方法。

由此可知，现有研究缺乏对 ADX 平台的总体研究。目前，无论是国内还是全球范围内，都有大量存在竞争关系的 ADX 平台。对于这些 ADX 平台，其广告投放有何特征以及投放效果如何，目前研究还处于一片空白。

考虑到以上问题，本文希望建立一个基于可靠数据、可扩展的跨 ADX 平台测量系统。该系统以多个 ADX 平台作为研究对象，对其行为定向广告投放效果做比较分析，衡量各个广告平台行为定向的能力大小、优势与劣势以及其他特点。

## 2.4　本章小结

本章介绍了在线广告系统的相关研究，主要介绍了在线广告生态系统的系统组成，包括系统架构、广告投放机制、定向广告和本文研究的行为定向广告，还介绍了在线广告系统所用的相关技术以及现有的研究内容，并总结了现有研究的优缺点。

# 3　同步跨平台测量方法

在本章，将详细介绍本文提出的同步跨平台测量方法。首先，介绍该方法的提出背景，使读者理解该方法的价值。接着，阐述该方法的基本思想，并开发设计一套实现该方法的系统。最后，详细介绍了系统各个模块的设计过程和编程实现。

## 3.1　跨平台测量

### 3.1.1　跨平台测量的提出

在线广告是互联网经济的主要动力，为各种网站和服务带来了可观收入。无论在工业界还是学术界，在线广告系统各参与者都对 ADX 有极大兴趣。对于研究者来说，ADX 平台就像一个黑盒子，为了研究 ADX 平台，网络测量是目前普遍采用的研究手段

（1）传统广告测量：

现有的测量是对单个平台进行测量，可以称之为传统广告测量模式。如图 3-1 所示，在线广告系统包括广告主、ADX 平台、媒体主和用户，广告主与 ADX1 合作，将自己要投放的广告交付给 ADX1。ADX1 在线追踪上网用户、挖掘其兴趣，并根据用户兴趣向用户投放相应的广告。

对单个平台进行研究，目前通常采用的测量方法是：先定义某个虚拟角色，在平台监测的网页中找到与虚拟角色特匹配的网页；通过多次访问这些网页，使该平台认识到有这个特定的虚拟角色存在；进一步，收集该平台针对该虚拟角色投放的广告，测量这些广告与该虚拟角色兴趣特征的相关性。采用构建虚拟角色的方法，而不考虑利用真实的用户进行实验，是因为人工操作任务量大且不可进行大规模实验。

（2）跨平台的广告测量

实际上，在线广告系统由很多存在竞争关系交易平台构成。在国外，主要有雅虎的 RightMedia，谷歌的 DoubleClick 等；在国内，百度、阿里 Tanx、新浪 SAX 等 ADX 平台成为主流平台。

在众多 ADX 平台中，如何选择 ADX 平台对在线广告系统的各个参与者都很重要。例如，图 3-2 所示的三个 ADX 平台存在竞争关系。广告主需要考虑选择哪一个交易平台进行合作，才能进一步提高他们的投资回报率；对于上网用户来说，

ADX 平台行为定向的效果会改善其上网体验感，所以媒体主需要考虑选择哪个 ADX 能吸引更多用户；对于 ADX 平台本身，优秀的广告投放性能可以吸引更多与其合作的对象，获得更多的收入。

因此，跨平台的广告测量是各方都亟需的现实需求。而现有研究是对单个广告交易平台进行测量的，无法满足这些需求。本文正是利用现有研究在这方面的空白，提出了跨平台的广告测量。

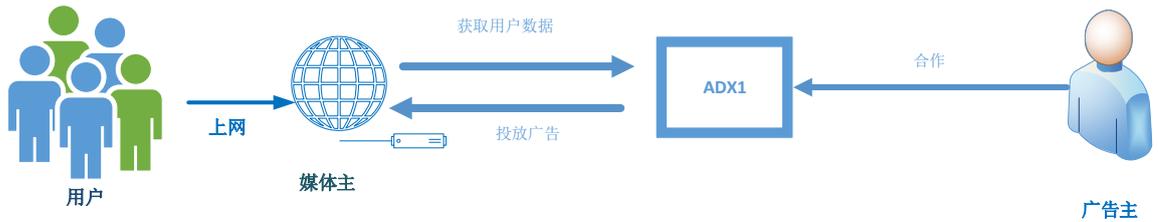

图 3-1 传统测量场景
Figure 3-1 Traditional measurement scenarios

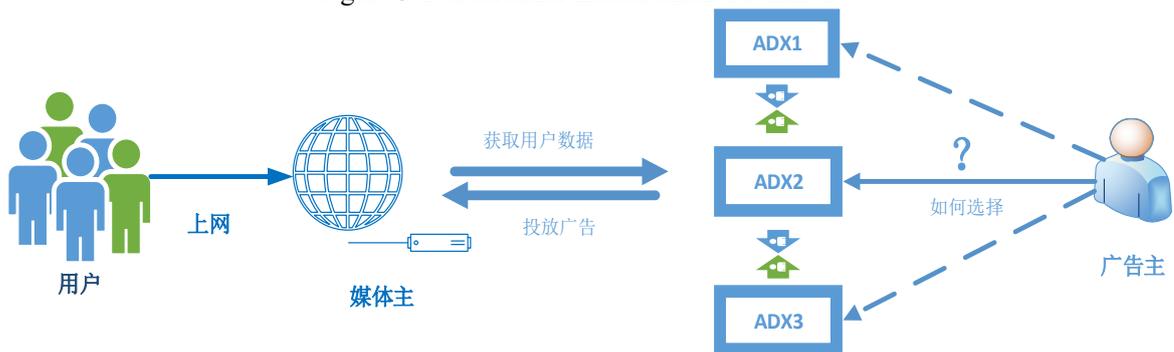

图 3-2 跨平台测量场景
Figure 3-2 Cross-platform measurement scenarios

### 3.1.2 跨平台测量的挑战

跨平台测量不能简单搬用传统的单平台测量方法，它将遇到如下新挑战：

挑战 1：跨平台测量不是多个单平台测量的简单重复或叠加。为了实现对多个平台的测量，一个朴素的想法是，用已有的单平台测量方法，逐个测量每个待测的平台。例如，为了比较谷歌和百度两个平台的广告投放效果，使用现有研究的方法对它们分别进行测量，再做比较分析。但是，这种想法实际上是不可行的。理由如下：

同时测量多平台需要有相同的测量基准，即需要在不同平台塑造相同的虚拟角色。如果采用单平台测量方法，通过逐个测量来研究多个平台，会存在以下问题：尽管单平台系统可以塑造兴趣特征相同的虚拟人，但是，这些系统单独塑造的虚拟人不可避免具有不同特征偏差。例如，在新浪平台和谷歌平台塑造一个吃货角色，

对新浪交易平台，假定可以通过访问新浪美食网页来塑造该角色；对谷歌交易平台，假定可以通过访问谷歌投放广告的美食网页来塑造该角色。但是，这两组网页之间是有差别的，新浪美食网页中可能还包含旅游的信息，谷歌美食网页中可能还有教育的信息。这会导致，新浪认为该角色除了喜欢美食之外还爱好旅行；谷歌认为该角色除了吃还关注教育。这样在两个平台中塑造的吃货角色并非完全一致，因此使用单平台测量方法逐个测量多平台并不准确。需要强调的一点是，本论文是对谷歌广告平台进行研究，而不是 google.com，测量研究中只需使用谷歌平台搜集其投放的广告。这是两个不同的系统，谷歌广告平台目前不受国家防火墙限制。

挑战 2：传统测量方法中广告提取方式并不适用于多平台比较测量。在传统的测量方法中，需要同时访问网页及网页中的广告内容，但是一旦读取广告内容，会造成平台对虚拟角色的认识偏差，为了解决这个问题，传统方法访问训练网页集合后再获取广告内容。此方法存在的问题是：由于获取广告链接和获取广告内容之间存在时间差，后续访问到的广告内容可能已经不是当时获取广告链接所对应的内容。因此但平台测量方法可能影响对平台广告投放效果的评估

挑战 3：传统测量方法识别广告内容的手段存在局限。在识别广告内容时，传统方法是将获取的广告链接提交给第三方网站，第三方网站返回该链接网页的所属类别。但是，采用这种方法，广告识别的准确度取决于第三方网站知识库的完备程度，并且还受制于第三方网站的使用权限。

## 3.2 同步跨平台测量方法

### 3.2.1 基本思想

为了解决上述挑战，本文提出"同步跨平台测量方法"，其基本思想如下：利用多个 ADX 平台共同监测的网页训练具有某种兴趣的虚拟角色，并获取不同 ADX 平台针对该角色投放的广告；对这些广告进行比较分析，从而评估和比较不同 ADX 平台的广告投放性能。

该方法具体如下：

第 1 步，找到这些平台监测的网页的交集。因为 ADX 在媒体主页面嵌入广告代码，使得他们可以监测到访问这些页面的用户的行为，并从中挖掘出用户兴趣。图 3-3 显示了一个用户访问的网页及网页关联的监测平台（广告平台），其中，ADX 平台 1 监测的网页范围是集合 1；平台 2 监测了集合 2 部分；平台 3 监测的是集合 3 中的网页；他们共同监测的网页是虚线框里的交集。本文找到这些交集的网页（具体方法见 3.3 节），并用作刻画虚拟角色的基本素材。

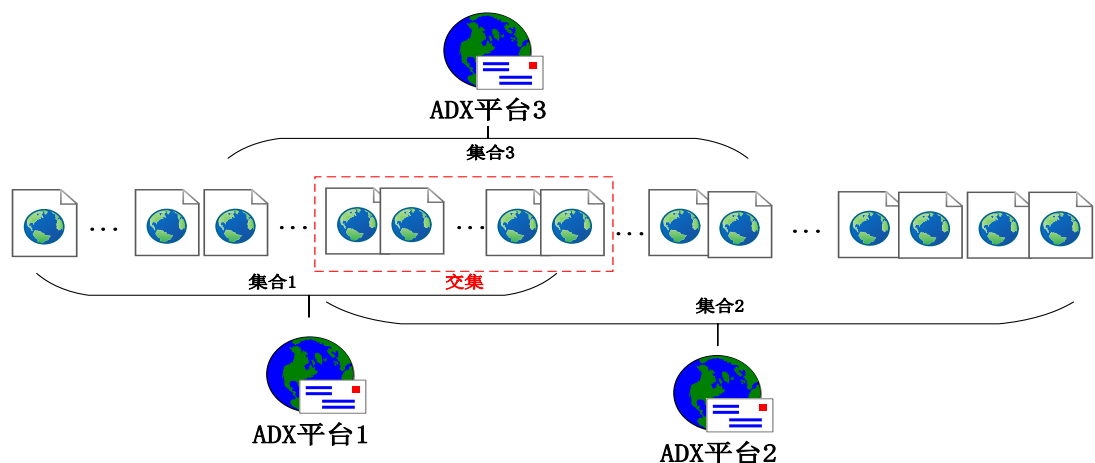

图 3-3 ADX 平台监测用户

Figure 3-3 ADX platform monitor users

第 2 步，构建基于兴趣的虚拟角色。对以上获得的交集网页进行分类，每个类别对应一种可能的虚拟角色。如果需要训练某个类别的虚拟人，就从该类别的网页中选取并访问 Top k 个网页(k 是设计参数，本文取 k=10)，由此在这些平台形成预期兴趣的虚拟角色。

第 3 步，在形成虚拟角色的过程中，本文还从这些交集网页中，选取并访问一组中性网页(不具兴趣偏好的网页)，获取这些网页上的广告链接。

第 4 步，基于上一步获得的广告链接，进一步获取广告内容。一方面，并行访问广告链接是本文提出的一个创新方法。在获取广告链接的同时用另一个专职读取广告的虚拟角色并发访问这些广告链接。采用这种方法以后，既能实时获取广告链接对应的广告内容，又不会造成对虚拟角色认识的偏差。另一方面，本文使用了基于图像识别的广告内容提取的方法。除了用文字识别，本文还使用了开放的图像检索引擎分析广告内容。这种方法在获得较好精度的同时摆脱了对第三方网站的依赖。这种方法得益于近年来图像识别和文本处理技术的发展。

第 5 步，基于获得的广告内容，分析不同广告平台的投放性能。

综上所述，本文提出的跨平台测量方法能够有效解决前文提出的挑战。该方法利用不同 ADX 的交集网页形成虚拟角色，解决了 3.2 节中指出的不同 ADX 对角色认识偏差问题。在 3.2 节挑战 1 的例子中，若要在谷歌和新浪平台同时构建一个吃货角色，不能采用新浪美食网页，因为这些网页不被谷歌平台监测；也不能采用谷歌美食网页，因为它不被新浪平台监测。该方法找到被新浪和谷歌平台同时监测的网页，当访问这些网页时，谷歌和新浪平台能够同时认识到吃货角色的存在。由于采用的是他们监测的交集网页，因此排除了平台对角色的认识偏差。

最后，在上述方法中，第 1 步测量交集的方法是本文针对跨平台测量提出的创新方法；第 2 步基于已有的网页进行虚拟角色训练和第 3 步采用中性网页手机广告都是广告测量研究中的常规方法；第 4 步多身份并行访问广告链接是本文提

出的创新方法；第 5 步采用了传统的性能分析方法。

### 3.2.2 系统框架

基于以上思想，本文设计了一套同步跨平台广告测量系统。该系统可实现三个主要功能：跨平台监测网页解析、基于兴趣的虚拟角色构建和广告识别。相应地，本文将通过设计三个模块来实现这些功能。系统整体框架如图 3-4 所示。

（1） 跨平台监测网页解析

跨平台监测网页解析是本系统要实现的首要任务，它将解决系统设计面临的最大挑战，也就是选取多个 ADX 平台共同监测的网页作为训练网页，为在多个 ADX 平台上构建相同兴趣的虚拟角色做准备。因此，跨平台监测网页解析模块的主要功能是广泛地采集网页，并解析各个 ADX 平台的监测网页。

（2） 基于兴趣的虚拟角色构建

获取了多个 ADX 平台共同监测的网页后，需要利用这些网页训练具有某种兴趣的虚拟角色。根据兴趣标签，可将解析模块获取的网页分为教育、旅游、体育等类别，选定一个兴趣类别的网页作为训练网页，并随机访问训练网页和控制网页的合集，可同时达到训练虚拟角色和收集广告的目的。

（3） 广告识别

广告识别是指识别各个 ADX 平台针对虚拟角色投放的广告具体是什么类别。根据广告样式的不同，该模块使用了三种方法对广告内容进行鉴别：基于 HTML 网页识别、基于图片识别以及基于关键词识别。

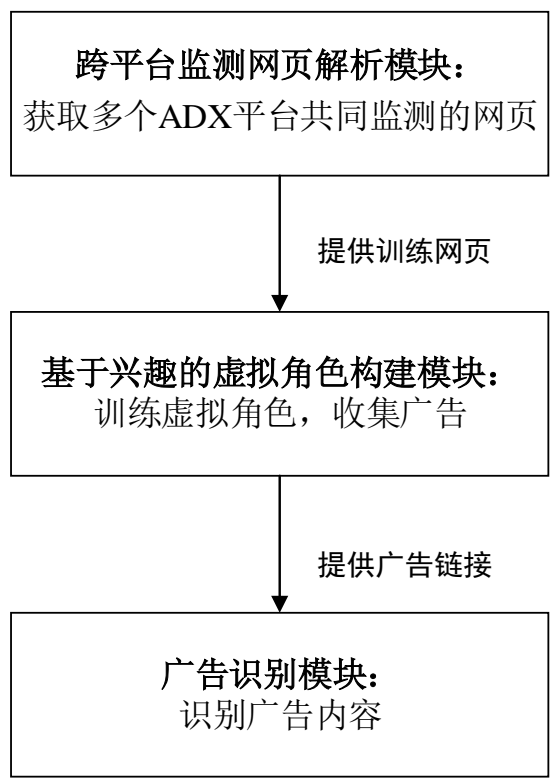

图 3-4 同步跨平台测量系统架构

Figure 3-4 Synchronous cross-platform measurement system architecture

接下来，本文将详细论述系统各个模块的设计思想、所用的方法技术以及相关细节。

## 3.3 跨平台监测网页解析

跨平台监测网页解析模块是本系统最基础的模块，该模块的主要目标是为训练网页提供必要的数据集。本小节将详细论述跨平台监测网页解析模块的设计思想和具体实现方法。

### 3.3.1 关键问题及其解决方法

为了对 ADX 平台的性能进行评估，需要在多个 ADX 系统中构建相同兴趣特征的虚拟角色。各个ADX平台由于监测网页不一样，获取的用户行为特征不一样，对用户兴趣的把握也不同。如果采用传统测量方法，这些平台单独塑造的虚拟人不可避免具有不同特征偏差。因此，只有利用它们共同监测的网页，才能在多个 ADX 系统中构建相同兴趣的虚拟角色。

本文提出了跨平台监测网页解析方法。该方法区别于传统测量方法之处在于，

它要找到多个平台监测网页的交集。如图 3-5 所示，要比较 ADX1 和 ADX2 两个平台的性能，必须先找到它们共同监测的网页，即蓝色部分，这些网页是刻画虚拟角色的基本素材。

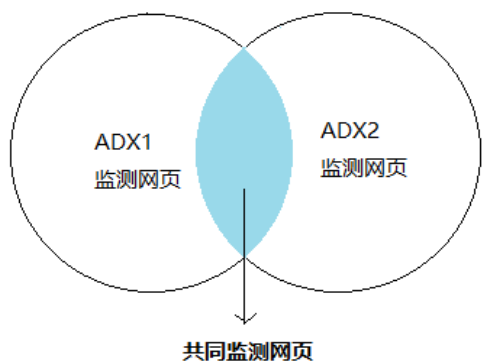

图 3-5 ADX 监测网页

Figure 3-5 ADX monitoring webpages

然而，要解决上述问题，面临的难点在于：（1）如何找到多个 ADX 平台共同监测的网页，（2）如何找到足够多的网页，用于刻画不同行为偏好的虚拟角色。为了获取满足以上要求的训练网页，本文尝试采取了两种方法。

一种可能的方法是参考知名的网站分类目录 DMOZ，从已经分好类的网站中挑选被 ADX 平台监测的网页。该方法提供已经分好类别的网页，可直接从这些网页中选取多个 ADX 平台共同监测的网页，训练相应兴趣特征的虚拟角色。但是经过实验发现，DMOZ 目录下的中文网页数量太少，完全不足以支撑大量训练网页的需求。

另一种方法是爬取大规模的网页数据，根据第三方追踪技术解析每个网页上的监测平台，找出被多个平台共同监测的网页。这种方法爬取的网页规模大，可提供较多交集网页；但是由于需要处理的数据量太大，花费的时间周期长。考虑到该方法的优点，本文最终采取这种方法来得到交集网页。按照这种方法：

首先，需要获取足够的数据集，为此需要广泛地爬取网页。一种可行的办法是，利用 Alexa 网站提供的大量网页排名数据，本文采用了全球流量排名前一百万的网页名单作为基本数据源。

其次，为了解析各个网页上的监测平台，需要找到网页和监测平台的关联。如何自动化、大规模地找出网页和监测平台间的关联是一个难题，为了解决该问题，本文利用第三方追踪技术的原理（详见 3.3.2 节），构建了一套映射准则。利用该准则，可以实现大规模网页自动解析，找出网页与 ADX 平台的对照关系。在此基础上，选取多个平台共同监测的网页作为后续构建虚拟角色的素材。

为了实现上述方法，本研究设计如图 3-6 所示的处理过程。

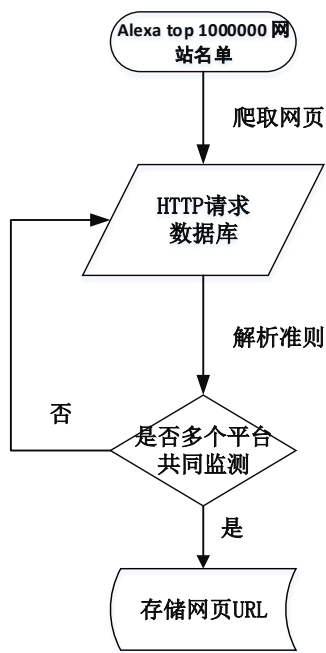

图 3-6 跨平台监测网页解析模块流程图

Figure 3-6 Cross-platform monitoring webpage parsing module flowchart

### 3.3.2 方案设计

（1）数据采集

为了研究谷歌和国内几大 ADX 平台（如：百度、阿里妈妈等）广告投放的性能，需要广泛收集它们投放广告的网页。为了获取一个较大的数据集进行研究，本文根据 Alexa 网站发布的网站世界排名，在排名前一百万的网站中进行寻找。

本文利用文献[61]中开发的 web 测量平台 OpenWPM 进行网络测量。它是一种开源网络隐私测量工具，不少现有研究都是在此基础上进行测量的。利用该工具，可以实现对大规模网站自动化访问，获取屏幕快照、访问日志等。将 Alexa 前一百万网站名单以列表形式传入 OpenWPM，可以获取到这一百万网站的 http request、http responses 等数据。

（2）解析 ADX 平台

解析网页上的监测平台是本模块的关键点。为了投放广告，ADX 平台在网页上嵌入广告代码，并对浏览该网页的用户进行监测。经观察发现，各个 ADX 平台

嵌入网页的广告代码分别指向一个或多个特定的源网址，这些源网址正是投放广告的平台。因此，本文构建了一套 ADX 平台及其源网址的映射准则，根据此准则，可以解析出网页上的监测平台。

按照本文构建的映射准则，对获取到的网站 HTTP 请求数据做分析，可以解析各网页上的 ADX 平台。具体原理如下：

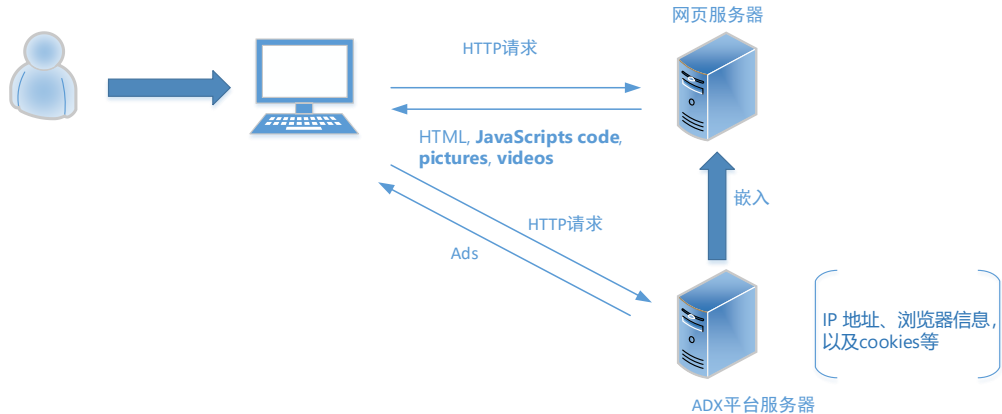

图 3-7 第三方追踪技术原理图

Figure 3-7 Third-party tracking technology diagram

如图 3-7 所示，当用户浏览一个网页时，其使用的浏览器向该网页的服务器发送了'GET'HTTP 请求，服务器将返回包含该网页 HTML 代码的 HTTP 响应。该 HTML 代码除了能使用户浏览器正常加载网页之外，通常还包含指向第三方服务器的样式表、JavaScript 代码以及图片。于是接下来用户浏览器会自动向这些第三方发送 HTTP request 请求，并且执行其代码。在网页上投放广告的 ADX 平台通常就是以第三方的角色存在，因此，通过分析访问网站时其发送的 HTTP request 请求数据可以识别出嵌入在该网页上的 ADX 平台。

数据采集部分获取的 HTTP request 数据中记录了 url、top_level_url、referrer 及 headers 等信息，如图 3-8 所示，top_level_url 是当前访问的网站，url 是该网站上的所有链接，referrer 是这些链接对应的源网址，可以从中分析出嵌在该网站上的第三方。

| id | crawl_id | visit_id | url | top_level_url | method | referrer | headers |
|---|---|---|---|---|---|---|---|
| 过滤 | 过滤 | 过滤 | 过滤 | 过滤 | 过滤 | 过滤 | 过滤 |
| 334 | 2 | 6 | https://pagead2.googlesyndication.com/pub-config/r20160913/ca-pub-8250252968172543.js | https://fone4u.com.pk/ | GET | https://fone4u.com.… | [["Host","p… |
| 335 | 2 | 6 | https://googleads.g.doubleclick.net/pagead/html/r20170724/r20170110/zrt_lookup.html# | https://fone4u.com.pk/ | GET | https://fone4u.com.… | [["Host","g… |
| 336 | 2 | 6 | https://pagead2.googlesyndication.com/pagead/js/r20170724/r20170110/show_ads_impl.js | https://fone4u.com.pk/ | GET | https://fone4u.com.… | [["Host","p… |

图 3-8 获取的 HTTP request 数据

Figure 3-8 HTTP request data

通过使用 Firefox 浏览器审查元素工具查看网站上的广告元素，可以发现各个 ADX 平台的广告总是指向特定的源网址。例如，Google 投放的广告链接都是以

http://googleads.g.doubleclick.net 开头的,而百度广告都是指向 https://cpro.baidu.com 或 http://pos.baidu.com。根据此发现，本文建立了一套 ADX 平台及其广告源地址的映射准则,表 3-1 给出了各个 ADX 平台及其广告源网页的映射关系。也就是说，如果网页的 HTTP 请求数据中含有这些广告源地址，则可以判断该网页被源地址对应的 ADX 监测。因此，对 HTTP request 数据进行正则匹配可以判断该网页上嵌入了哪些 ADX 平台。

表 3-1 各 ADX 平台广告源地址

Table 3-1 ADX platform ad source address

| ADX 平台 | 广告源地址 |
| --- | --- |
| 谷歌 | https://googleads.g.doubleclick.net、https://ad.doubleclick.net |
| 百度 | http://pos.baidu.com 、https://cpro.baidu.com、http://baidustatic.com |
| 阿里 | http://atanx.alicdn.com、http://click.tanx.com |
| 苏宁 | http://wmt.qtmojo.com |
| 京东 | https://ccc-x.jd.com/dsp/、http://ads-union.jd.com |
| 新浪 | http://sax.sina.com |
| 搜狗 | http://union.sogou.com、http://galaxy.sogoucdn.com |

由于 ADX 平台数量较多，本文选择几个主流的 ADX 平台作为研究对象，包括谷歌、百度、阿里、苏宁和京东等。图 3-9 显示了部分网页上的监测平台，"1"表示该网页上包含对应的平台，"0"表示该网页不被对应的平台监测。对于同时包含多个平台的网页，将其挑选出来作为训练网页。

| | | baidu | taobao | google | suning | sina | jd | sogou |
| --- | --- | --- | --- | --- | --- | --- | --- | --- |
| 2 | http://www.forever21.cn/ | 1 | 0 | 1 | 0 | 0 | 0 | 0 |
| 3 | http://www.renren.com/ | 1 | 0 | 0 | 0 | 0 | 0 | 0 |
| 4 | http://www.mydrivers.com/ | 1 | 0 | 0 | 0 | 0 | 0 | 0 |
| 5 | http://www.iciba.com/ | 1 | 0 | 1 | 0 | 0 | 0 | 0 |
| 6 | http://www.zhibo8.cc/index.html | 1 | 0 | 0 | 0 | 0 | 0 | 0 |
| 7 | http://www.zol.com.cn/ | 1 | 0 | 1 | 0 | 0 | 0 | 0 |
| 8 | http://www.nipic.com/index.html | 1 | 0 | 0 | 0 | 0 | 0 | 0 |
| 9 | http://www.taoche.com/ | 1 | 0 | 1 | 0 | 0 | 0 | 0 |
| 10 | http://www.3dmgame.com/ | 1 | 0 | 0 | 0 | 0 | 0 | 0 |
| 11 | https://hypebeast.cn/ | 1 | 0 | 1 | 0 | 0 | 0 | 0 |
| 12 | http://365jia.cn/ | 1 | 0 | 0 | 0 | 0 | 0 | 0 |
| 13 | http://www.5dcar.com/ | 1 | 0 | 0 | 0 | 0 | 0 | 0 |
| 14 | http://www.hjenglish.com/ | 1 | 0 | 0 | 0 | 0 | 0 | 0 |
| 15 | http://www.dji.com/cn | 1 | 0 | 1 | 1 | 1 | 0 | 0 |
| 16 | http://www.baimao.com/ | 1 | 0 | 0 | 0 | 0 | 0 | 0 |
| 17 | http://beijing.anjuke.com/ | 1 | 0 | 1 | 0 | 0 | 0 | 0 |
| 18 | https://www.ihg.com/hotels/cn/zh/reservation | 1 | 0 | 1 | 0 | 0 | 0 | 0 |

图 3-9 网页上的 ADX 平台

Figure 3-9 ADX platform on the web

对于挑选出的训练网页，使用本人设计的网络爬虫获取网页内容，为之后网页分类做准备。但要注意的是，爬取的网页对象是全球流量排名前一百万的网站，这其中有大部分是非中文网站，而谷歌在这些非中文网站上也会投放广告。为了和国内 ADX 平台做比较，需过滤掉非中文网站。

## 3.4 基于兴趣的虚拟角色构建

基于兴趣的虚拟角色构建模块的主要目标是：训练产生具有某种兴趣特征的虚拟角色并收集不同 ADX 针对该角色投放的广告。本小节将第一部分阐述了该模块的设计思想，后面两部分详细论述了网页分类、训练虚拟角色及收集广告的具体方法。

### 3.4.1 关键问题及其解决方法

该过程希望创建一些虚拟角色，并呈现与某特定兴趣相对应的网页浏览行为，例如"运动"或"烹饪"。 通过访问精心挑选的符合其兴趣的网页来培训每个虚拟角色，也可以使广告交易平台相应地对虚拟角色进行分类。将访问过的网站称为训练网页。例如，"赛车运动"角色的训练集将由类别为赛车运动的网页构成。

因此，有三个关键问题需要考虑：挑选哪些虚拟角色来研究、在哪些网站上收集广告以及怎么排除其它类型广告（与角色偏好无关）。

对于虚拟角色的选择，需要同时考虑到代表性和实际性。现有的研究对不同兴趣用户给行为定向广告造成的影响做了测量，而本文提出的方法侧重于比较多个 ADX 平台，因此需要在各种兴趣中挑选具有代表性又满足广告投放实际情况的。由文献[52]可知，不同兴趣的用户对定向广告的影响不同，本文研究的虚拟角色应该囊括各种影响度的兴趣，这样才能保证对广告系统的全面研究。而在实际应用中，训练网页是由多个 ADX 共同监测网页组成的，根据真实用户的兴趣标签体系，可以将这些网页进行分类，并得到多个兴趣类别（如：教育、旅游、金融等）。本文需要考虑这些网页的类别，在已有类别中挑选虚拟角色的兴趣。

一旦选定虚拟角色和相应的训练网页，还需要收集各个 ADX 平台针对虚拟角色定向投放的广告。收集广告的网页被称为控制网页，它需要满足以下标准：1）页面上有足够多的展示广告，且这些广告属于多个 ADX 平台；2）页面内容没有明显兴趣偏向或非常好定义，可以很容易地检测到基于上下文的广告，并将其过滤出来，只保留那些基于用户兴趣的广告。为此，本文选择门户网站的页面作为控制网页。

我们想要获取的广告是 ADX 平台针对虚拟角色兴趣投放的定向广告，而广告系统中除了此类型广告外，还包含其他类型的广告（详见 2.1 节）。因此为了排除广告系统固有的广告投放内容带来的偏差，还需要考虑构建一个不具有兴趣偏好的空白角色。

## 3.4.2 方案设计

### 3.4.2.1 网页分类

在上一小节获取了多个 ADX 平台共同监测的网页,对这些网页分类,每个类别可作为一种虚拟角色的训练网页。训练网页的类别直接影响虚拟角色的兴趣特征,也就是说,网页分类越精准,虚拟角色的兴趣特征越明显。

根据网页文本的特性,对网页分类的处理过程包括:文本预处理、文本表示、聚类和归纳。文本表示和聚类可统称为网页聚类,为了达到更好的效果,在此部分利用两种方法进行对比实验,并最终采用聚类效果较好的方法。

接下来将介绍整个网页分类的处理过程。

(1) 文本预处理

网页文本的内容是用自然语言描述的,是计算机不能理解的,因此必须先将文本转化成计算机可处理的数据。文本预处理是文本表示的基本前提。文本预处理包括:去除数据中非文本部分以及中文分词两个过程。

针对通过网络爬虫获取的网页数据,去除数据中非文本部分主要指将 HTML 的标签去掉。如果爬取的数据中只含有少量非中文内容,可以直接用 Python 中的正则表达式(re)去除;如果数据中含有的非中文部分太复杂,可以使用 BeautifulSoup 来去除。

中文分词是文本预处理最重要的过程,在上一小节中过滤掉非中文网站,获取到的都是中文网页数据。中文文本中词与词之间没有明显的切分标志,所以需要分词处理,提取能代表文档的字词或词组让计算机理解文档的主要意思。本文使用最常用的结巴分词工具对中文文本进行分词,分词后每个文档将由多个词条构成。

(2) 网页聚类

1) 基于 Doc2vec 和 Agglomerative Clustering 的网页聚类

Doc2vec 是基于 Word2vec 的基础上发展而来的方法,它可以将一段句子表征为实数值向量。通过 Doc2vec 将网页文本表示成向量后,采用 Agglomerative Clustering 算法对网页聚类。Agglomerative Clustering 是一种自底向上的层次聚类方法,可以设置 3 种不同的聚类规则:Complete(maximum) linkage,两类间的距离用最远点距离表示;Average linkage,平均距离;Ward's method,以组内平方和最小,组间平方和最大为目的。

2) 基于 kMeans 和 TF-IDF 的网页分类

在多种文本表示模型(如布尔模型、聚类模型、向量空间模型概率模型、基于知识的表示模型以及混合模型等)中,向量空间模型是近年来应用较多、效果较好

的一种，其基本思想是将文档表示为加权的特征向量：D=D(T1,W1;T2,W2;…;Tn,Wn)。其中 Tn 为特征项词条，Wn 为特征词条的权重。

经过文本预处理后，文档的词条可能会达到数千维甚至上万维，本文从中选择一部分具有特征值的词条作为特征项，选取的标准大多是以词频统计为基础。最后对所有的词条按照其权值的大小排序，选取预定数目的最佳特征作为文本的表示向量，本文选取了前 20 个词作为特征项。

权值计算公式是 $tf_{i,j}*idf$，它是一种用于资讯检索与文本挖掘的常用加权技术。TF-IDF 是一种统计方法，用以评估一字词对于一个文件集或一个语料库中的其中一份文件的重要程度。字词的重要性随着它在文件中出现的次数成正比增加，但同时会随着它在语料库中出现的频率成反比下降。其标准表达式为[62]：

$$tfidf_{i,j} = tf_{i,j} * idf_i \tag{3-1}$$

其中，

$$tf_{i,j} = \frac{n_{i,j}}{\sum_k n_{k,j}} \tag{3-2}$$

$$idf_i = \log_2 \frac{|D|}{|\{j: t_i \in d_j\}|} \tag{3-3}$$

以上式子中 $n_{i,j}$ 是该词 $t_i$ 在文件 $d_j$ 中的出现次数，而分母则是在文件中 $d_j$ 所有字词的出现次数之和。$|D|$：语料库中的文件总数，$|\{j: t_i \in d_j\}|$ 包含词语 $t_i$ 的文件数目（即 $n_{i,j} \neq 0$ 的文件数目）

向量空间模型将每个网页文本表示成一个向量，而所有的网页文本则组成一个 N*M 矩阵，其中 N 为网页文本个数，M 为特征词数。

利用向量空间表示网页文本后，采用 KMeans 算法对网页聚类。KMeans 聚类是一种无监督学习，它接受输入量 k，然后将数据对象划分为 k 个聚类并使得：同一聚类中的对象相似度较高，而不同聚类中的对象相似度较小。kMeans 的聚类效果受输入量 k 值影响，而 k 值是由用户给定的。

3）聚类优化

无论是使用 Agglomerative Clustering 层次聚类还是 KMeans 聚类，聚类个数 k 都是由用户给定的。为了确定合适的 k 值使聚类效果最好，引入 Silhouette Coefficient 作为评估指标。

轮廓系数（Silhouette Coefficient）适用于实际类别信息未知的情况。对于单个样本，设 a 是与它同类别其他样本的平均距离，b 是与它距离最近不同类别中样本的平均距离，轮廓系数为[63]：

$$s = \frac{b - a}{\max(a, b)} \tag{3-4}$$

对于一个集合样本,它的轮廓系数是所有样本轮廓系数的平均值 $\overline{S}(k)$。轮廓系数取值范围是[-1,1],值越接近 1,聚类效果越好。

因此,利用最优公式(3-5),能够找出使这个指标分数最高的 k 值,从而达到最好的聚类效果。

$$\begin{aligned} \max &: \overline{S}(k) \\ s.t. &\ 1 \leq k \leq m \end{aligned} \tag{3-5}$$

采用上述两种聚类方法对京东平台的监测网页聚类,图 3-10(a)图和(b)图分别显示了这两种方法选定不同 k 值时的轮廓系数。图中的横坐标表示聚类个数 k 的取值,这里的范围都是 0~30,纵坐标表示轮廓系数值。(a)图所示的是基于 Doc2vec 的聚类方法轮廓系数与 k 值关系,在 k=25 时轮廓系数最大;(b)图显示的是基于 kMeans 的聚类方法轮廓系数与 k 值关系,在 k=4 时轮廓系数最大。由于两种方法给出的聚类效果最好时的 k 值差别较大,并且第一种方法轮廓系数最大只达到 0.021,第二种方法轮廓系数最大可达到 0.35,因此本文选择采用聚类效果更好的第二种方法,也就是基于 kMeans 的聚类方法。经过人工验证,最终发现基于 kMeans 的聚类方法聚类效果的确比较好

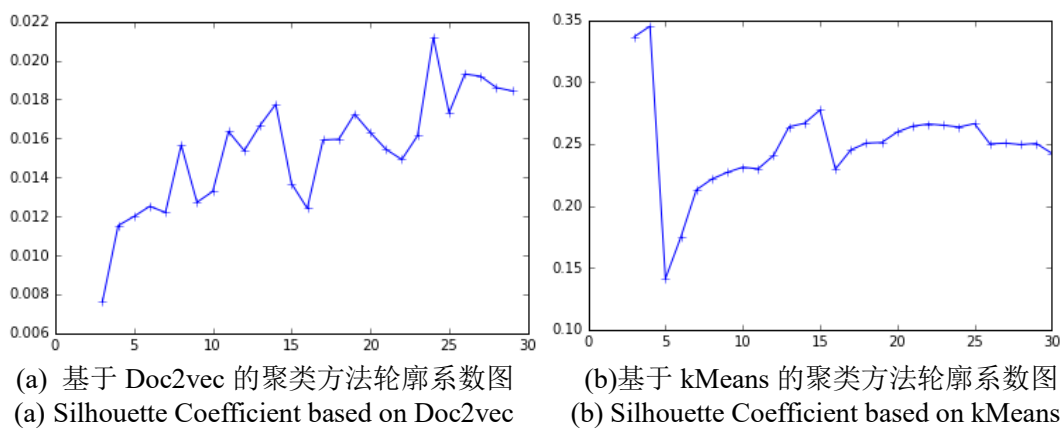

(a) 基于 Doc2vec 的聚类方法轮廓系数图    (b)基于 kMeans 的聚类方法轮廓系数图
(a) Silhouette Coefficient based on Doc2vec    (b) Silhouette Coefficient based on kMeans

图 3-10 轮廓系数图

Figure 3-10 Silhouette Coefficient

4) 归纳

聚类后相似的网页文本归入一簇,还需要为其定义类别。为了方便接下来对用户兴趣标签和广告标签进行评估,在这里采用同一标签体系为用户兴趣和广告划分类别。Google AdWords 体系是 Google 为个性化广告定制的分类结构,如表 3-2 所示,它将广告分为艺术、运动等 32 个大类。每个大类又细分为多个小类,表 3-3 介绍了最大的三个类别下的子类别。根据每簇的前几个关键词,可以将这些簇划

分到 Google 广告体系类别中。由此，可以给聚类后的网页给定相应的主题，即类别名。

表 3-2 Google AdWords 广告类别

Table 3-2 Google AdWords ad category

| 艺术 | 娱乐 | 计算/技术 | 政治 |
|---|---|---|---|
| 农业 | 教育 | 住宅 | 房地产 |
| 动物 | 家庭/育儿 | 法律 | 宗教 |
| 建筑 | 时尚 | 军事 | 科学 |
| 成人 | 民俗 | 新闻 | 社会 |
| 汽车 | 食物/饮料 | 金融 | 体育 |
| 商业 | 健康/健身 | 宠物 | 购物 |
| 职业 | 历史 | 哲学 | 旅游 |

表 3-3 三大类别的子类

Table 3-3 Subcategories of the three categories

| 顶级类别 | 子类别 |
|---|---|
| 艺术 | 设计、美术、文学、摄影、音乐、歌剧、诗歌、小说 |
| 健康/健身 | 感冒/流感、哮喘、解剖、自闭症、心脏病、聋哑、口腔医学、皮肤医学、糖尿病、节食、癫痫、运动、眼部护理、急救、药品、精神病、营养学、外科、儿科、理疗、心理学、保健、失眠、戒烟、维生素、非传统医学 |
| 金融 | 银行储蓄、贷款、信用卡、财务计划、保险、投资、养老、股票、纳税 |

### 3.4.2.2　训练虚拟角色及收集广告

该部分的主要目标是：利用已经分类的网页，挑选不同类别的网页用于训练相应兴趣的虚拟角色并收集广告。

最简单的训练虚拟角色的方法是设置客户端访问训练网页。设置客户端随机访问由训练网页和控制网页组成的集合，可同时达到训练虚拟角色和收集广告的效果。通过修改客户端 user-agent，可以构造多种虚拟角色，因此该系统具有自动化和可扩展性。

多身份并行访问广告链接是本文提出的一个改进的广告提取方法。在提取广告时，传统的方法是先访问所有网页序列获取广告链接集合，然后再获取广告链接中对应的内容。此方法存在的问题是：由于获取广告链接和获取广告内容之间存在

时间差，后续访问到的广告内容可能已经不是当时获取广告链接所对应的内容。而本文在获取广告链接的同时用另一台机器并发访问这些广告链接，另一台机器具有与原来的机器不同的身份。采用这种方法以后，既能实时获取广告链接对应的广告内容，又不会造成对虚拟角色认识的偏差。

要实现该目标，主要包括三个步骤：虚拟角色和训练网页的选择、控制网页的选择，以及训练虚拟角色并收集广告。

（1）虚拟角色和训练网页的选择

Google Adwords 类别体系提供的兴趣标签有几十上百种，选择哪些虚拟角色来进行研究是一个难题。既希望虚拟角色的行为偏好比较鲜明，又不愿研究对象过于特例化而失掉总体性，但构造太多的虚拟角色会给本研究带来繁重的工作量，因此参考文献[52]中给出的不同兴趣用户对行为定向的影响程度，如表 3-4 所示，分别选择对行为定向影响较大、适中以及较小的 1~2 个虚拟角色进行研究。

表 3-4 兴趣对行为定向影响程度

Table 3-4 Impact on OBA

| 影响程度 | 兴趣标签 |
| --- | --- |
| 较大 | 教育/语言学习，旅行，游泳，自行车/配件，会计/审计，社交网络，金融，露营，汽车保养 |
| 适中 | 视觉艺术/设计，宠物产品，编程，园艺，电影，地理参考，文件共享，美发，琐事/趣闻，天文学 |
| 较小 | 古玩/收藏，诗歌，军事，烹饪，网络小说，娱乐新闻，论坛 |

（2）控制页面的选择

一定选定虚拟角色和训练网页，需要收集各个 ADX 平台针对虚拟角色定向投放的广告。收集广告的网页被称为控制网页，选择门户网站的页面作为控制网页。因为门户网站页面上有足够多不同 ADX 平台的展示广告，并且页面内容没有明显兴趣偏向或非常好定义。

（3）训练虚拟角色并收集广告

一旦为一个角色选择了一套训练网页和控制页面，将按照以下策略访问它们。如图 3-11 所示，把 10 个训练网页和 5 个控制页面放到一个列表中，按一定时间间隔从列表中随机选择一个页面进行访问，访问时间间隔服从均值为 3 分钟的指数分布。采取这种方法的原因有两个：一方面，可以定期访问训练页面，以便这些网页上的 ADX 平台能根据浏览行为，对虚拟角色兴趣做出判断；另一方面，通过定期访问控制页面，可以收集向虚拟角色展示的广告，以便后续研究是否由行为定向所驱动。

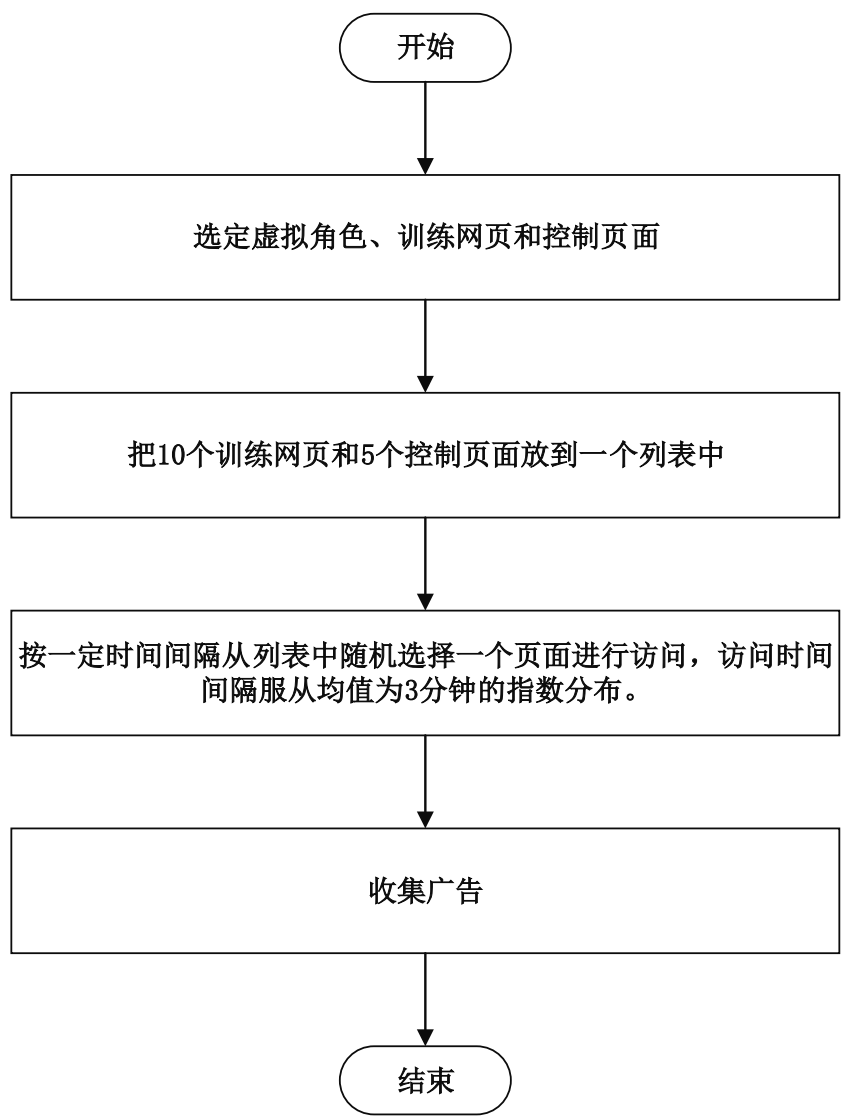

图 3-11 基于兴趣的虚拟角色训练流程图

Figure 3-11 The workflow diagram of training artificial online personas

除此之外，还有一种策略是首先多次访问训练页面，让广告平台记住虚拟角色，然后访问控制页面收集广告。本文并没有采用这个策略，因为连续多次访问控制页面会使得广告平台认为该的浏览器偏好"门户网站"。最后，为了避免在访问时点击广告给虚拟角色带来的角色污染，在整个实验结束之后再点击查看广告页面。

利用工具 Selenium 来获取广告，它是一个 Web 应用程序的软件测试框架，能够以编程的方式控制一个真正的网络浏览器（本实验中使用的 Mozilla Firefox）。这种方法可以检索呈现的广告的全部内容，如果使用像 wget 这样简单的命令行工具，这是不可能的。

为了保证实验的准确性，应使不同虚拟角色处于相同的广告市场。通过设置不同 user-agent 的客户端，系统在一段时间内同时训练多个虚拟角色并收集广告。此外，为了排除 ADX 系统固有广告的影响，构建一个空白角色，即没有兴趣偏好的

虚拟角色作为参照。表 3-5 展示了用于训练部分虚拟角色的客户端 user-agent。

表 3-5 虚拟角色用户代理

Table 3-5 Virtual role user agent

| 虚拟角色兴趣 | user-agent |
|---|---|
| 金融 | "Mozilla/5.4 (Macintosh; Intel Mac OS X) Chrome/56.0" |
| 设计 | "Mozilla/5.3 (X13; Linux x86_64) Firefox/55.2" |
| 军事 | "Mozilla/5.0 (Linux; Android 5.1.1) Safari/537.36" |
| 空白 | "Mozilla/5.0 (compatible; MSIE 9.0;Windows NT 6.1;Trident/5.0)" |

值得注意的是，在大多数情况下，广告被包含在 iframe 中。iframe 是嵌在 HTML 文档中的另一个 HTML 文档，即使它被不同的网站所包含，iframe 中的内容仍可以以一致的方式呈现。因此，本文抓取了所有 iframe 里的广告，包括静态和动态广告。

## 3.5 广告识别

提取广告落地页的关键字也可称为广告识别，本小节主要介绍本文使用的基于图像识别的广告内容提取方法。本文的目标是评估 ADX 平台广告投放的性能，也就是分析具有某种行为特征的用户与向广告平台向他们展示的广告之间的联系。

### 3.5.1 关键问题及其解决方法

广告识别是一个复杂的过程。在上一小节中，已经获取了控制网页产生的广告以及其链接，相应的可访问该链接到达广告落地页。对于这些广告页面，需要判断该广告商品属于哪个类别，也就需要提取广告落地页关键词。提取广告落地页的关键字也可称为广告识别。

传统测量方法在识别广告内容时，是将获取的广告链接提交给第三方网站，例如 Google，McAfee 和 Cyren[52]，第三方网站返回该链接网页的关键词。但是，采用这种方法，广告识别的准确度取决于第三方网站知识库的完备程度，例如在对中文网页进行标记时准确度很低甚至无法识别；此外，还受制于第三方网站的使用权限，例如第三方网站会设置验证码来防止大规模自动测量。

在传统方法基础上，本文使用了基于图像识别的广告内容提取的方法。此方法将直接获取广告链接对应的页面，然后用文字或图像识别技术分析广告内容。这种方法得益于近年来图像识别和文本处理技术的发展。这种方法摆脱了对第三方网站的依赖。

为了实现该方法，主要操作流程如图 3-12 所示。访问广告链接到达广告落地

页，首先判断网页主体是否为图片，若是则利用图像识别进行识别；若不是则进一步获取 HTML，利用标题或商品详情进行识别；此外，可以利用第三方工具获取该网页的关键词从而进行识别。

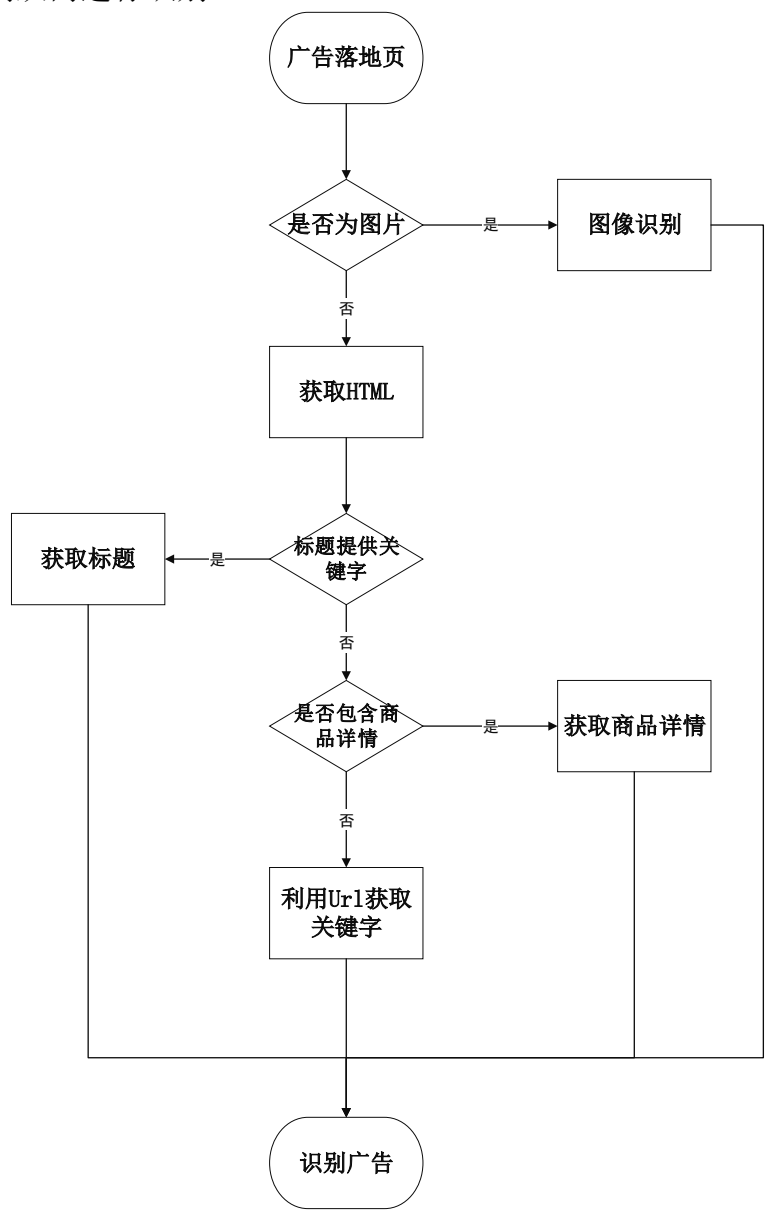

图 3-12  广告识别流程图

Figure 3-12 The workflow diagram of advertising recognition

### 3.5.2　方案设计

接下来将分别介绍该方法中所用的技术，包括基于 html 的广告识别、基于图片的广告识别、基于关键词的广告识别。

（1）　基于 HTML 的广告识别

利用的广告页面的 HTML 文档内容可判断广告内容。

网页文档作为互联网信息的一种载体，可以向访问网页的用户反馈各种各样的信息。HTML（超文本标记语言——HyperText Markup Language）是构成 Web 世界的基石。HTML 用于创建在万维网上显示的电子文档（称为页面）。每个页面都包含一系列连接到其他页面的超链接，在 Internet 上看到的每个网页都是使用某个版本 HTML 代码编写的。HTML 代码可以确保网页正确的文本和图像格式，以便用户的 Internet 浏览器可以按照他们打算查看的方式显示它们。如果没有 HTML，浏览器将不知道如何显示文本或加载图像或其他元素。HTML 也提供了页面的基本结构，通过层叠样式表可以改变其外观。

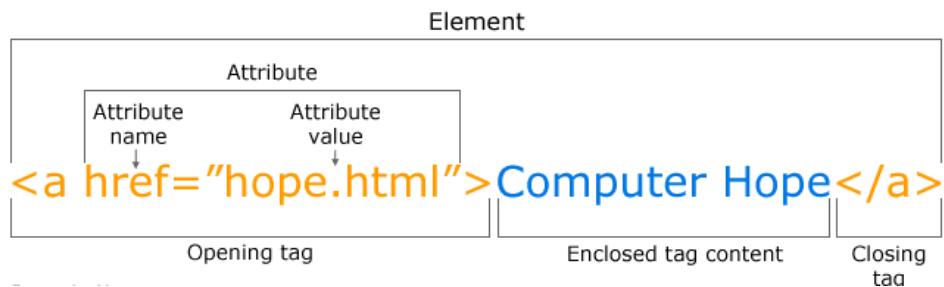

图 3-13 HTML 标记示例

Figure 3-13 HTML markup example

通过分析 HTML 的样式和构成，可以从海量网页数据中快速准确地获取自己所需要的信息。HTML 使用"标记"来注明文本、图片和其他内容，以便于在 Web 浏览器中显示。HTML 标记包含一些特殊"元素"如<head>、<title>、<body>、<p>、<div>和等等。如下面的 HTML 标记示例所示，组件数量不多。每个 HTML 标签都包含在小于和大于尖括号内，开始和结束标签之间的所有内容都被标签显示或影响。在图 3-13 中，<a>标签正在创建一个名为"Computer Hope"的指向 hope.html 文件的链接。接下来将介绍两种通过分析广告页面 HTML 标记内容来识别广告的方法：通过标题识别广告以及通过商品详情识别广告。

1) 通过网页标题识别广告

网页标题是一篇网页所要表达信息的最简明扼要的概述，它经常被应用于网页信息的处理及搜索引擎、网页聚类和分类等。

网页标题抽取方法有很多，常用的方法主要集中于基于机器学习和基于规则两类。本文使用的是基于正则表达式的网页标题提取方法，先通过 python 工具包 urllib 将网页内容抓取下来，然后用正则表达式 re 模块将标题匹配出来。

如图 3-14 所示，访问获取到的广告链接，进入其广告落地页后抓取其 HTML 文档中的<title>和</title>标签内的标题，可快速判断出这是一个智能触屏手表的广告。

```
<title>智能触屏手表手机支持微信qq</title>
```

图 3-14 HTML 中的标题

Figure 3-14 The title in the HTML

2） 通过商品详情识别广告

大多数情况下可以通过 HTML 文档中的<title>和</title>标签准确的获得"真实标题"，但很多时候网页的标题是比较笼统的描述，标题内容不能准确的指明此文档中的主要内容，或者只是一个域或网站的名字。就如下面的例子，图 3-15 中，广告落地页的广告商品是旗袍连衣裙，但其网页标题是"淘宝热卖"，该标题比较广泛而不能反映广告商品具体内容，因此可以采用获取商品详情的方式来判断广告商品。

```
▼<div class="details">
  ▼<div class="basic">
    ▼<h2>
        <a href="http://click.simba.taobao.com/cc_im?p=&s=776610039&k=416&e=SuvMGoiP%2Bk…
        lk1=64c597b4eaba0f748cb7bb052da4e294&clk1=64c597b4eaba0f748cb7bb052da4e294" target="_blank">金丝绒旗袍连
        衣裙</a>
      </h2>
    ▶<div class="basicContainer">…</div>
    </div>
  ▶<div class="similarItem">…</div> == $0
</div>
```

图 3-15 HTML 中的商品详情

Figure 3-15 Product details in HTML

一般来说，属于同一广告平台的广告网页结构基本相似，或干脆使用同一套网页模板，因此可以通过分析每个 ADX 平台的广告网页结构手工制定规则来获取广告商品的详情。

（2） 基于图片的广告识别

除了前文提到的具有网页标题或商品详情的广告之外，还有一部分广告网页展示的只有一张图片。在这种情况下，无论是根据 HTML 文档内容还是 URL 的关键词都无法判断出广告内容，因此采用图像识别技术来处理该部分广告。所谓图像识别技术，简单的来说，就是计算机通过对图像进行特定处理，理解其内容，从而找到用户所需要的信息。这些信息可能是一段文字，也可能是一段视频。在本文中，需要的信息是图片里的广告产品到底是什么。

图像识别是人工智能的一个重要领域，学术界对此有许多前沿的研究著作，而在实际中也出现了不少成熟的工具。为了减少在此部分的时间投入，选择 Google 图片的搜索功能作为工具，即利用 Google 的图像识别技术为广告图片贴标签，从而判断广告商品的内容。

如下图所示，尽管广告图片中没有出现直接描述广告产品的文字，Google 图片反馈的标签却准确地判断出该商品为佛珠。

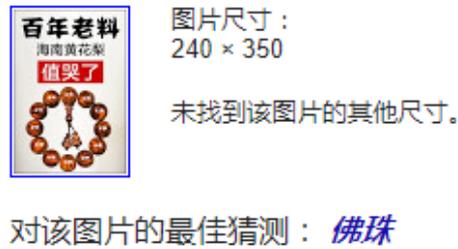

图 3-16 Google 图片识别

Figure 3-16 Google image recognition

（3） 基于关键词的广告识别

从上一小节可知一个标准规范的 HTML 文档应该在<title>…</title>标签中明确的标明网页文档的标题。但是，实际上并非所有网页的编辑者都能很准确的将文档的标题写在 Title 标签中，比如网上存在很多网页的 Title 标签中没有写任何信息。对于不能从标题中获取信息也没有商品详情的广告页面，可以利用 Google Adwords 关键字规划师工具获取其反馈的该网页关键词作为判定广告内容的参考。

Google AdWords 为其用户提供了一些帮助他们投放广告的工具，关键字规划师工具就是其中一种，它有两个主要功能：关键字提示和流量估算值。本文使用了其"网站相关关键词"选项，输入获取到的广告 URL，AdWords 将返回该广告页面上的不同关键词，并按照相似度分组。通常 Adwords 返回的前 5 个关键词就能描述出该广告商品的具体内容。

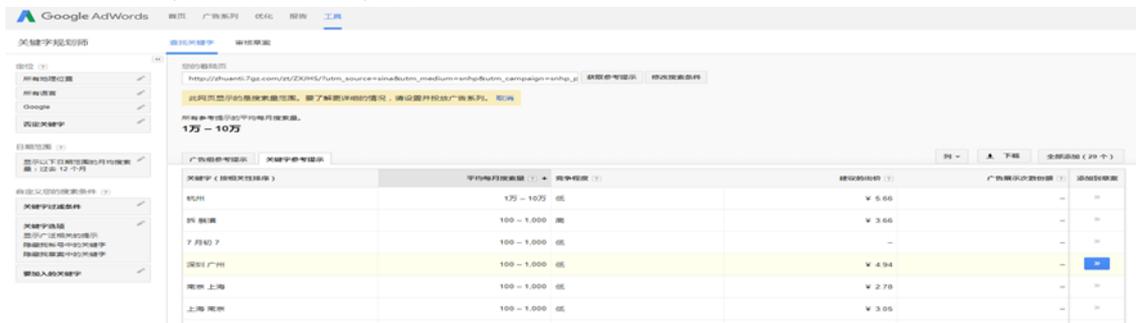

图 3-17 Google Adwords 关键字规划师工具

Figure 3-17 Google adwords keyword planner tool

上图就是一个装修广告的网页，在网页标题不能反映其真实内容的情况下，将网页 URL 送至 AdWords 的关键字工具，可以看到反馈的前 5 个关键词能够较准确地反映出这是一个装修广告。

## 3.6 编程实现

针对本文设计的同步跨平台测量系统，通过编程实现系统各个模块的功能。

为了实现跨平台网页监测解析模块功能，使用 OpenWPM 0.7.3 版本来驱动

Firefox 对 AlexaTop 一百万网页进行测量，并记录这些网页的所有 HTTP 数据。经过两个多月的测量，一共采集到了一百万网站的 59,307,628 条 HTTP 请求记录，解析出 10355 个含有研究平台广告的网页。该过程时间周期较长，数据处理量大。

为了实现基于兴趣的虚拟角色构建模块功能，使用了一个轻量级的无头浏览器 PhantomJS ver. 1.9[64]作为基础，通过修改客户端 user-agent，可以构造多种虚拟角色。该过程是在 Spyder 开发环境下编写 Python 语言脚本实现的，由于该过程是对多个 ADX 平台和多种虚拟角色处理，任务量较大，代码超过一千行，且训练时间持续了一个多月。

为了实现广告识别功能，使用了图像识别工具和自然语言处理技术，自主开发了基于图像和文字的广告识别脚本。

## 3.7　本章小结

本章详细介绍了本文提出的同步跨平台测量方法，包括该方法的提出、基本思想，以及为了实现该方法设计的系统。然后介绍了为了实现系统各功能，如何设计各个模块。最后，对设计的模块进行编程实现。

# 4 性能评估

为了评估本文设计开发的同步跨平台测量系统的有效性，利用该系统对实际的 ADX 平台进行了测量和比较分析。实验结果达到了预期效果，在同一基准下，该系统能够明显区分出不同ADX平台对相同虚拟角色的广告投放差异；并发现不同ADX平台广告投放会随虚拟角色改变而变化，且这种变化的灵敏程度存在不同。

本章首先介绍了实现测量的实验环境，包括实验的软硬件环境及实验步骤；然后给出了实验结果以及对结果的分析，包括对测量数据集的统计情况、评估指标的介绍和不同 ADX 平台的对比分析。

## 4.1 实验环境

### 4.1.1 软硬件环境

本文利用同步跨平台测量系统对多个 ADX 平台进行测量时，数据量比较大，操作较复杂。本文对硬件环境和软件环境的要求都比较高。因此，本文使用了高性能的电脑实现测量分析。

本文使用了两台电脑，配置如下：

电脑 1

处理器： Intel(R) Core(TM) i7-4590k CPU @ 4.00GHZ 4.00GHZ

内存（ RAM）： 16.00GB

系统类型： 64 位 windows 操作系统

电脑 2

处理器： Intel(R) Core(TM) i5-4590 CPU @ 3.30GHZ 3.30GHZ

内存（ RAM）： 8.00GB

系统类型： Ubuntu 14.04 操作系统

本文对网络测量环境有一定要求，将两台电脑接入校园网进行测量。

### 4.1.2 实验步骤

根据本文提出的同步跨平台测量方法，对百度、谷歌等多个 ADX 平台进行了测量。

首先，本文对 Alexa 前一百万网站进行测量，并解析各个平台的监测网页。

接着，利用多个平台共同监测的网页，训练几组不同兴趣的虚拟角色并收集平台针对这些角色投放的广告。

最后，识别收集到的广告，评估及比较不同 ADX 平台广告投放的效果。

## 4.2　实验结果分析

本小节介绍了对多个平台测量的结果，并对结果进行分析，评估及比较了多个平台的广告投放性能。第一部分介绍了测量数据的整体统计情况，包括各个 ADX 平台监测网页及交集网页数，以及投放广告的网页类别。第二部分根据广告的统计结果对两组平台进行比较分析。第三部分分析了不同 ADX 平台的时间动态行为特征。

### 4.2.1　测量数据集的整体统计

经过测量，共获取到一万多个监测网页。各个 ADX 平台网页监测情况如图 4-1 所示，可观察到，百度和谷歌平台投放广告的数量遥遥领先其他平台，占了整个广告投放网页的 91%，这说明百度和谷歌广告平台投放广告的范围广、数量多，在广泛投放广告方面有优势。此外，阿里和苏宁平台的监测网页数也达到数百个，属于比较流行的广告平台。腾讯、网易等平台的监测网页数量较少。一个猜测是这些平台可能由于自身特性，倾向于在主题明显但流量较小的网站上投放广告，这些网站不在流量排名前一百万的名单上，无法分析。

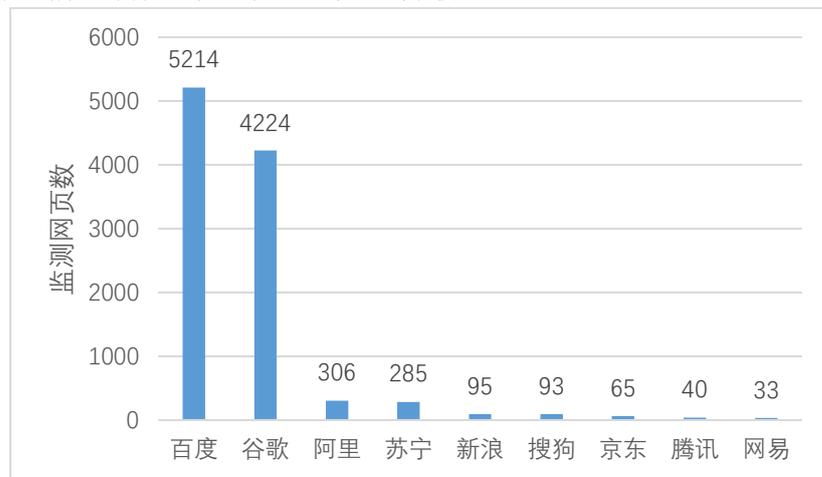

图 4-1 各 ADX 平台投放广告网页数

Figure 4-1 The number of pages served on each ADX platform

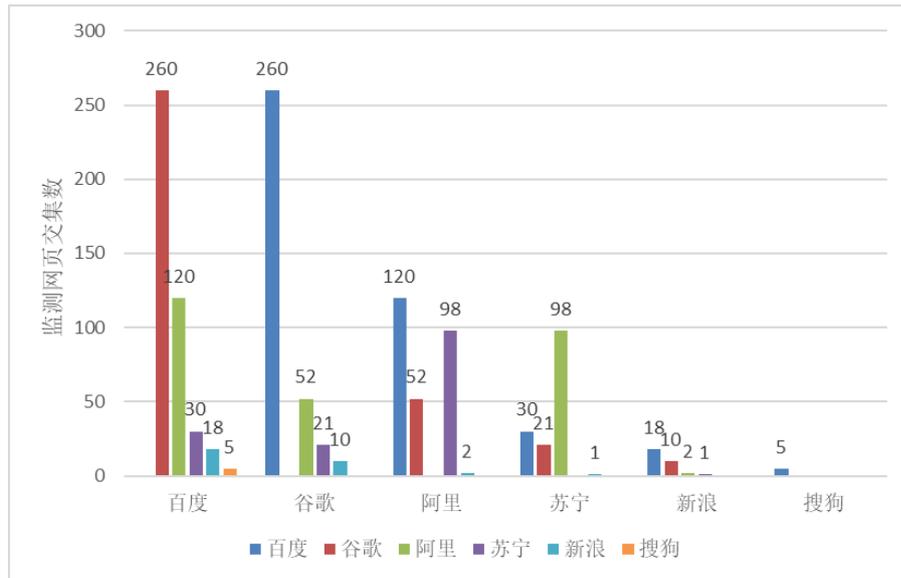

图 4-2 各 ADX 平台监测网页类别

Figure 4-2 ADX platform monitoring pages category

利用本文提出的方法，找到各个 ADX 平台的监测网页交集。由于三个及三个以上平台共同监测的网页数太少，本文只分析了两个平台共同监测网页的情况。两个平台监测网页的交集情况如图 4-2 所示，横坐标表示百度等 6 个 ADX 平台，不同颜色的条形代表与该平台有共同监测网页的其它平台以及交集网页数目。百度和谷歌由于监测网页数较多，他们共同监测的网页数是最多的，有 260 个；百度与阿里的监测网页交集有 120 个；阿里与苏宁的监测网页交集有 98 个；其余的平台之间监测网页交集数都较少。

根据两个 ADX 平台共同监测网页的情况，本文选择百度与谷歌、阿里与苏宁两组平台分别进行评估及比较。原因如下：它们的监测网页交集比较多，足够用于训练多种虚拟角色；百度和谷歌都具有搜索引擎的功能，阿里和苏宁都是电商平台，两组研究平台功能相似，有比较研究的意义。

通过对这些监测网页进行分类，可总结出各个ADX平台投放广告的类别偏好。图 4-3 展示了百度、谷歌两个平台监测网页的类别以及他们共同监测网页的类别。由图可知，百度和谷歌平台投放广告的网页类别占比最多的都是计算/技术类，分别达到了 21%和 26%。百度平台在新闻、教育、娱乐和艺术等类别网站都有较多的广告投放，且涉及到的网站类别有 18 个之多，基本涵盖了用户上网浏览的所有网站类别。相比之下，谷歌广告交易平台投放广告的网页类别只有 9 类，且主要集中在购物、教育、旅游和金融领域。百度和谷歌共同监测的网页共有 9 类，其中占比最多的是计算/技术、教育和金融类网页，因此在选择虚拟角色时需要从这些占比较多的监测网页交集中选取。

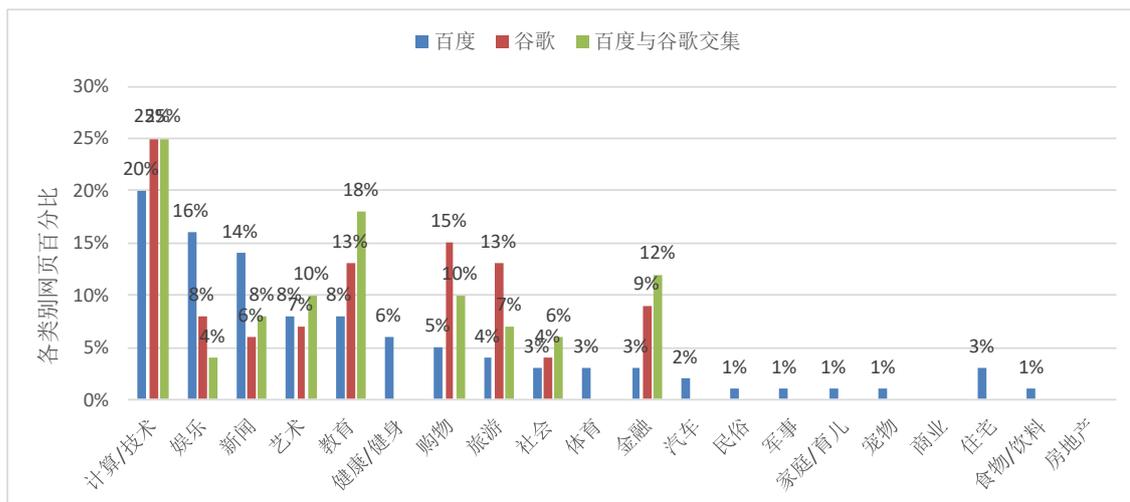

图 4-3 百度与谷歌监测网页类别

Figure 4-3 Baidu and Google monitoring pages categories

图 4-4 显示了阿里与苏宁平台监测网页的类别以及他们共同监测网页的类别。阿里平台在艺术、新闻和娱乐等类别网页投放广告较多，苏宁平台则倾向于商业、购物和住宅等类别的网页，他们共同监测的网页主要集中在娱乐、教育和艺术等领域。

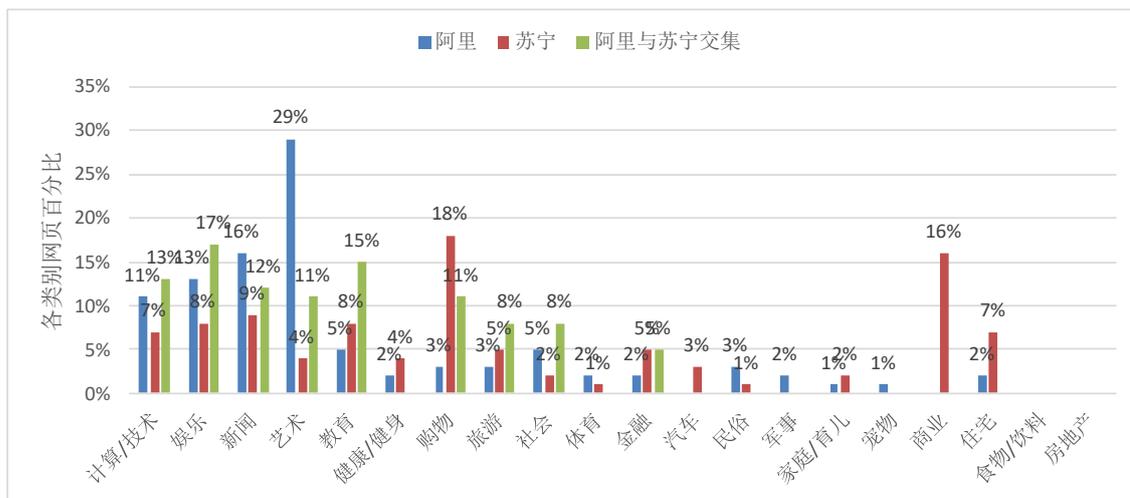

图 4-4 阿里与谷歌监测网页类别

Figure 4-4 Ali and Suning monitoring pages categories

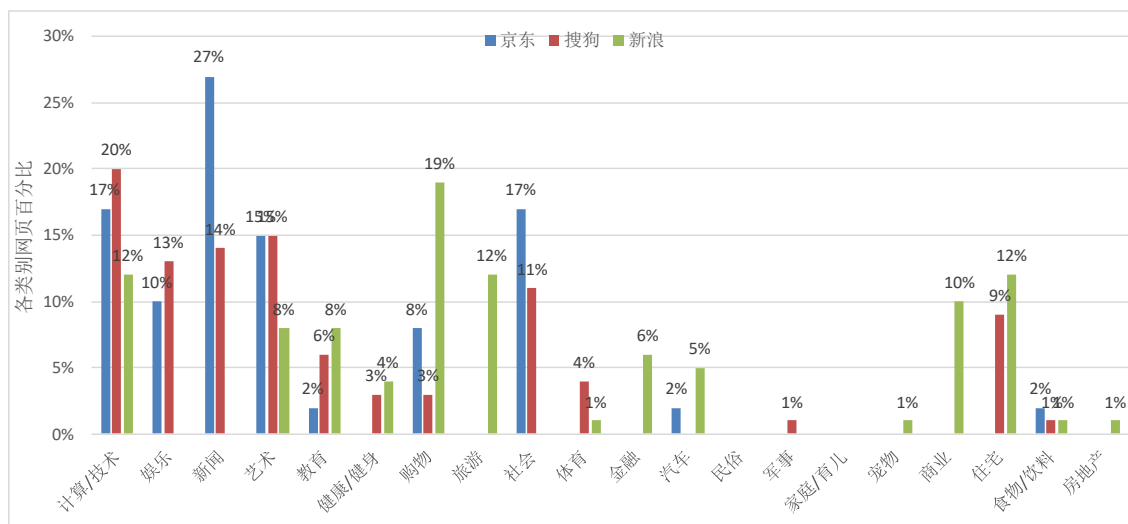

图 4-5 京东、搜狗及新浪监测网页类别

Figure 4-5 JD、Sougou and Sina monitoring pages categories

对于京东、新浪和搜狗平台获取到的广告投放网页数量不多，但也简单地统计了其分布。如图 4-5 所示，京东平台主要在新闻、计算/技术和社会等类别网页投放广告；新浪平台主要在购物、计算/技术、旅游和住宅类别的网站投放广告，搜狗平台的广告主要投放于计算/技术、和新闻网站上。

## 4.2.2　评估指标介绍

本文的目标是评估及比较 ADX 平台广告投放的性能，也就是分析具有某种兴趣特征的虚拟角色与广告平台向他们展示的广告之间的联系。

为了实现该目标，本文使用简单的、易于理解的衡量指标，利用虚拟角色的兴趣标签和广告落地页的关键字来量化这种性能。其具体方法如下。

对于具有兴趣 i 的虚拟角色，通过计算该角色训练网页和广告落地页关键词的重叠关系，可以衡量 ADX 平台广告投放性能。具体而言，本文使用两个互补的度量标准来衡量训练网页和广告落地页关键字重叠的不同方面。在此之前，先介绍一下指标中使用的一些定义：（1）网页分类时提取了各网页关键词，对于训练具有兴趣 i 的虚拟角色的训练网页，称其关键词集合为 $K_{T_i}$，表 4-1 展示了部分兴趣标签所对应的关键词集合；（2）对于虚拟角色收到的广告，其广告落地页关键词集合为 $K_{L_i}$；（3）称空白角色收到的广告落地页关键词集合为 $K_k$，该集合是不受虚拟角色兴趣影响的。利用这些定义，采用两种评估指标[52]：

表 4-1 部分兴趣标签关键词

Table 4-1 Part of interest tags keywords

| 兴趣标签 | 关键词 |
| --- | --- |
| 体育 | 篮球、足球、兵乓球、网球、排球、羽毛球、跑步、游泳、滑冰 |
| 音乐 | 歌曲、交响乐、吉他、口琴、钢琴、风琴、小提琴、铜管乐器 |
| 金融 | 金融、投资、股票、信托、证券、贷款、保险 |
| 计算/技术 | IT、API、云计算、编程、算法、人工智能、大数据 |

训练网页关键词定向（TTK）：该指标计算了训练网页关键词被定向的比例，也就是计算训练网页关键词集合中有多少关键词呈现在广告落地页上。其具体表达式如下：

$$TTK(i) = \frac{\left| K_{T_i} \cap (K_{L_i} - K_k) \right|}{\left| K_{T_i} \right|} \in [0, 1] \quad (4\text{-}1)$$

实质上，TTK 测量虚拟角色的兴趣 i 是否会产生行为定向。具体来说，TTK 的值比较大表示具有兴趣 i 的虚拟角色会收到 ADX 平台的定向广告。

落地页行为定向（BAiLP）：该评估指标捕获了广告落地页关键词集合与训练网页关键词的交集。当空白角色收到的广告包含训练网页关键词时，需要先排除这部分广告的影响。换句话说，BAiLP 代表了由虚拟角色兴趣 i 带来的行为定向。BAiLP 正式表述如下：

$$BAiLP(i) = \frac{f(K_{L_i})}{K_{L_i}} \in [0, 1]$$

$$其中\ f(K_{L_i}) = \begin{cases} (K_{L_i} - K_k) \cap K_{T_i}, & K_{T_i} \cap K_k \geq 1 \\ K_{L_i} \cap K_{T_i}, & K_{T_i} \cap K_k = 0 \end{cases} \quad (4\text{-}2)$$

总之，TTK 测量由兴趣 i 产生行为定向广告的概率，而 BAiLP 则衡量虚拟角色收到的广告中行为定向广告的比例。

## 4.2.3 不同 ADX 平台对比测量及分析

对于百度和谷歌、阿里与苏宁两组平台，分别进行对比测量并评估他们的广告投放效果。根据他们监测网页交集的情况，在百度与谷歌平台上，选定"金融"、"设计"和"军事"三种虚拟角色进行测量比较；在阿里和苏宁平台上，选定"教育"、"电影"和"小说"三种角色进行研究。

### 4.2.3.1　百度与谷歌

百度和谷歌作为搜索引擎，都拥有较大的互联网流量和用户数据信息，因此将对这两个 ADX 平台进行比较，评估它们针对"金融"、"设计"和"军事"三种虚拟角色的定向广告投放效果。

利用同时包含百度和谷歌广告的网页训练产生具有金融、设计和军事偏好的虚拟角色，对于训练好的三种虚拟角色，在门户网站上分别收集每种角色下展示的百度广告和谷歌广告，并统计各个类别广告所占百分比。通过与空白角色获取到的广告内容做对比，可以分析出两个 ADX 平台针对每种虚拟角色定向投放广告的程度。

图 4-6 和图 4-7 分别展示了谷歌和百度平台针对不同虚拟角色投放的广告类别及占比。图中四种颜色分别代表空白角色、金融、设计和军事爱好者，折线上的点表示每个类型的广告出现的比例。对于不同角色，百度平台和谷歌平台投放的广告类型不一样且差别较大。针对"金融"和"设计"角色，两个平台投放广告都发生了较大变化，而对"军事"角色却都没有定向投放广告。产生这一现象的原因可能有两种：广告平台对这类角色的定向能力比较弱；军事类型的广告本身就很少。而究竟是由第一种还是第二种或者两种原因同时造成，这需要作另外的研究才可知。

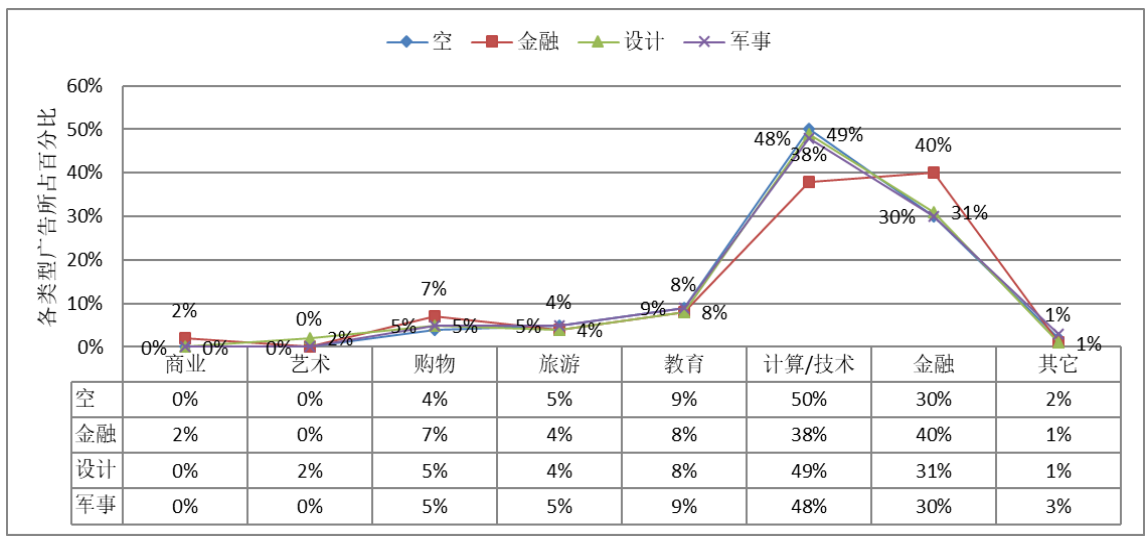

图 4-6 谷歌 ADX 平台行为定向广告效果图

Figure 4-6 Google ADX platform behavioral targeted advertising effectiveness

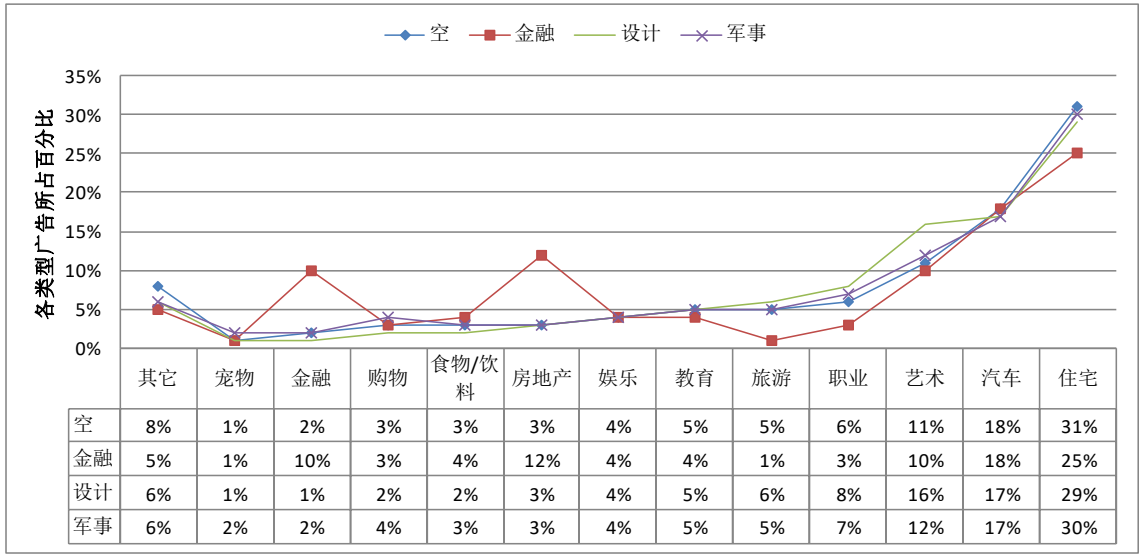

图 4-7 百度 ADX 平台行为定向广告效果图

Figure 4-7 Baidu ADX platform behavioral targeted advertising effectiveness

为了对比谷歌平台和百度平台的投放效果，比较分析了这两个平台针对同一种角色投放广告百分比及改变程度。图 4-8 和图 4-9 分别是两个平台针对"金融"和"设计"角色投放广告的情况。

由图 4-8 可知，针对"金融"类角色，百度平台下金融类别广告从 2%增长到 10%，增加了 8%。除此之外，房地产类别的广告从 3%增加到 12%，也有大幅度的增长，这可或许是因为百度广告平台认为房地产和金融相关度较大。谷歌平台在空白角色下金融类别广告已经达到 30%，在虚拟角色影响下增长了 10%，有明显的增幅。

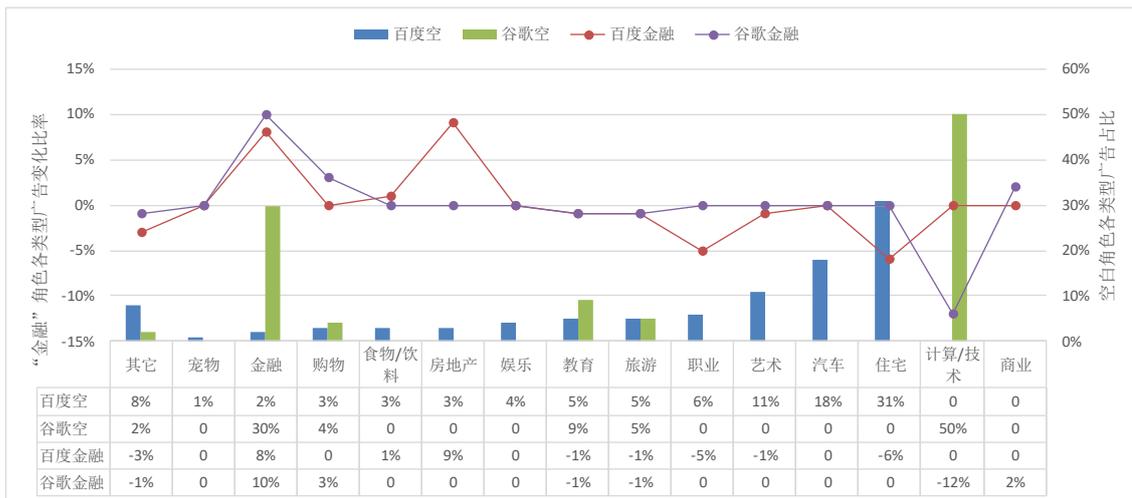

图 4-8 "金融"角色收到的广告

Figure 4-8 Advertising received by "finance" persona

由图 4-9 可知，针对"设计"角色，百度平台在"设计"虚拟角色影响下，艺术类广告从 11%增长到 16%；谷歌平台艺术类别广告从零增长到 2%，增幅不大，

但实现了从无到有的过程。

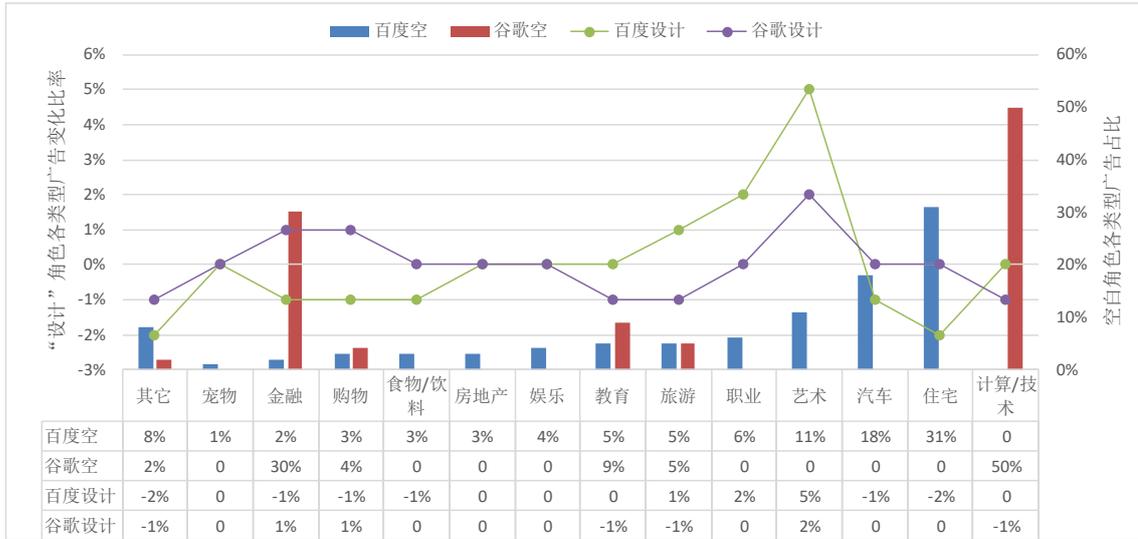

图 4-9 "设计"角色收到的广告

Figure 4-9 Advertising received by "design" persona

根据以上分析，初步可得出谷歌和百度平台会针对虚拟角色兴趣定向投放广告的结论。接下来，将利用指标 TTK 进行进一步评估。

TTK 测量虚拟角色的兴趣 i 是否会产生行为定向。具体来说，TTK 的值比较大表示具有兴趣 i 的虚拟角色会收到 ADX 平台的定向广告。

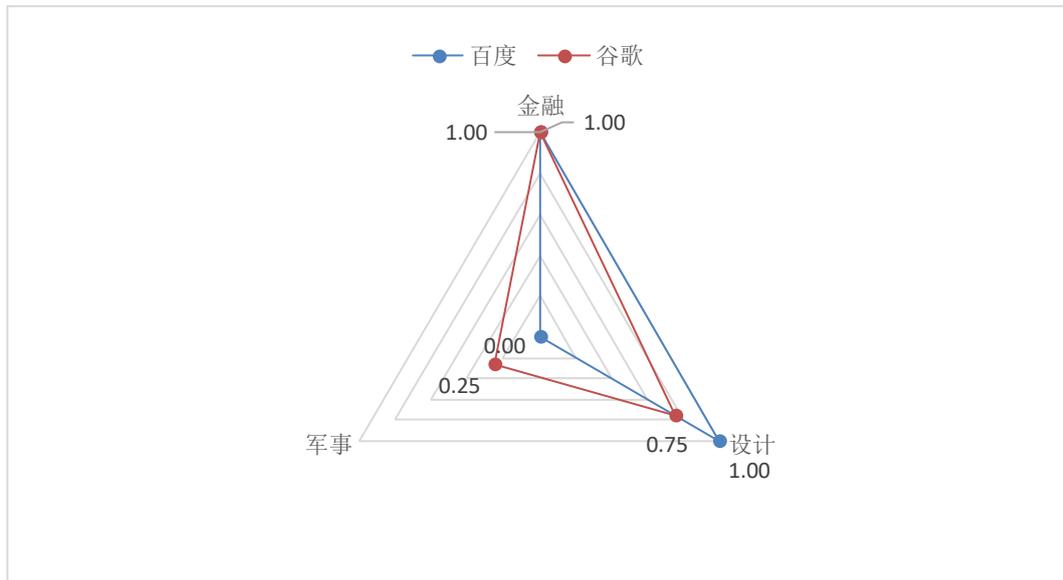

图 4-10 百度与谷歌 TTK 值

Figure 4-10 TTK of Baidu and Google

图 4-10 是谷歌和百度的 TTK 值，反映了这两个平台下的三种角色是否能产生行为定向广告。由图可知，百度和谷歌平台针对"金融"和"设计"虚拟角色都能产生行为定向广告，谷歌对"军事"角色也产生了行为定向广告，这在之前的结果中没有显示，可能是统计占比情况下该类广告太少被忽略了。

总的来说，无论是百度还是谷歌广告交易平台，它们都会针对用户行为偏好定

向投放广告。随着用户偏好不同，百度定向广告投放最多可增加 8%，谷歌最多增长 10%。虽然百度平台增幅稍微比谷歌平台小，但其同时会在和用户偏好相关度较大的类别增加广告投放，也就是说百度除了能针对用户现有偏好投放广告还可以智能联想用户的其它兴趣领域。此外，正如文献[52]中用户偏好对定向广告影响大小的结论，偏好金融的用户行为对定向广告影响最大，设计类对定向广告产生有影响但改变不多，军事类几乎没有影响。

#### 4.2.3.2 阿里与苏宁

就大众所熟知，淘宝和苏宁易购是电商行业的两大巨头，它们都希望售卖更多的商品来获取利润，而广告作为一个促进利益增长的重要方式对它们来说尤为重要。因此本文希望对阿里的 Tanx 广告平台和苏宁的易通天下广告平台进行对比分析，评估它们针对用户行为偏好定向投放广告的效果。

根据阿里和苏宁投放广告的网页类别分布，选择"教育"、"电影"和"小说"三种用户偏好进行研究。通过分析收集到的广告，可以发现这两个广告平台投放的广告内容都是其电商平台上的商品，并且呈现出多且繁杂的特点。因此，在这里并没有统计各个类型广告所占的比例，而是以是否出现和虚拟角色类型相契合的广告作为衡量广告平台定向广告投放效果的标准。也就是说，如果广告平台针对某一角色投放相应的广告，那可以判定其行为定向能力较强。

表 4-1 至表 4-3 分别呈现了一周内阿里广告平台向教育、电影和小说三种角色展示的广告类别。从表 4-1 可以看出，对于教育类的角色，从第三天起阿里平台开始对其展示童装、儿童电脑以及学生外套、钱包等广告，这些广告商品都明显包含"儿童"、"学生"等与教育密切相关的关键词。对于偏好电影的用户，可以从表 4-2 发现，在第三天其收到了懒人电脑桌、电动升降桌以及苏宁电视家影专场的广告。虽然这些广告商品并不包含电影这个关键词，但是它们完全就是为用户在家观看电影量身打造的。至于阿里广告平台会投放苏宁的广告，是因为阿里入资苏宁与其建立合作的关系。对于偏好小说的虚拟角色，并没有在阿里投放的广告里找到相关的商品，这可能是因为淘宝广告商品里没有小说这一类别。

表 4-1 "教育"角色收到的阿里 ADX 广告

Table 4-1 Advertising received by " education " role from ali ADX

| 阿里教育 | 广告落地页关键词 |
|---|---|
| Day1 | 旅游、家具、办公、冬装、装饰、运动、饰品 |
| Day2 | 旅游、家具、冬装、装饰、运动、饰品、保健品、保温杯、手电筒 |
| Day3 | 旅游、家具、冬装、装饰、饰品、茶壶、行李箱、童装 |
| Day4 | 家具、办公、冬装、运动、茶壶、童装、宝宝磨牙棒、儿童电脑 |
| Day5 | 办公、童装、儿童电脑、学生外套、零食、鞋 |
| Day6 | 冬装、饰品、童装、盆栽、玩具书本收纳箱、学生钱包 |
| Day7 | 家具、冬装、饰品、鞋 |

表 4-2 "电影"角色收到的阿里 ADX 广告

Table 4-2 Advertising received by " movie " role from ali ADX

| 阿里电影 | 广告落地页关键词 |
|---|---|
| Day1 | 冬装、家具、美妆、盆栽、旅游、户外、首饰 |
| Day2 | 冬装、包、家具、电脑、摆件 |
| Day3 | 冬装、首饰、耳机、懒人电脑桌、电动升降桌、苏宁（电视家影专场）|
| Day4 | 冬装、家具、包、首饰、盆栽、鞋 |
| Day5 | 冬装、家具、鞋、首饰、钓竿 |
| Day6 | 冬装、首饰、服务器、主板、路由 |
| Day7 | 冬装、钓竿、首饰、旅游、盆栽 |

表 4-3 "小说"角色收到的阿里 ADX 广告

Table 4-3 Advertising received by " novel " role from ali ADX

| 阿里小说 | 广告落地页关键词 |
|---|---|
| Day1 | 首饰、冬装、鞋、盆栽、家居、工艺品 |
| Day2 | 家居、首饰、盆栽、杯子 |
| Day3 | 冬装、钱包、首饰、茶壶、摆件 |
| Day4 | 冬装、盆栽、摆件、首饰、浮漂 |
| Day5 | 家具、摆件、冬装、工艺品 |
| Day6 | 冬装、鞋子、首饰、钓竿 |
| Day7 | 家具、首饰、盆栽、户外 |

表显 4-4 示了苏宁广告平台一周内的广告投放情况，在这里只有一张表的原因是实验中没有发现苏宁广告平台有针对用户偏好进行定向投放。正如表 4-4 所示，苏宁投放的广告基本以家电为主，这也与它主要业务为大家电的情况相符。

表 4-4 苏宁 ADX 平台广告

Table 4-4 Suning ADX platform advertising

| 苏宁 | 广告落地页关键词 |
|---|---|
| Day1 | 手机、冰箱、电视、热水器 |
| Day2 | 电磁炉、冰箱、平板电脑、移动硬盘 |
| Day3 | 手机、空调、电饭煲、热水器 |
| Day4 | 手机、相机、耳机、面包机 |
| Day5 | 冰箱、加湿器、热水壶、平板电脑 |
| Day6 | 空调、沙发、洗衣机、耳机 |
| Day7 | 汽车用品、平板电脑、电暖器、热水器 |

图 4-11 是阿里和苏宁平台的 TTK 值，反映了这两个平台下的三种角色是否能产生行为定向广告。如图中所示，阿里平台可以针对"教育"和"电影"角色定向投放广告，苏宁平台对三种角色都没有定向投放广告。

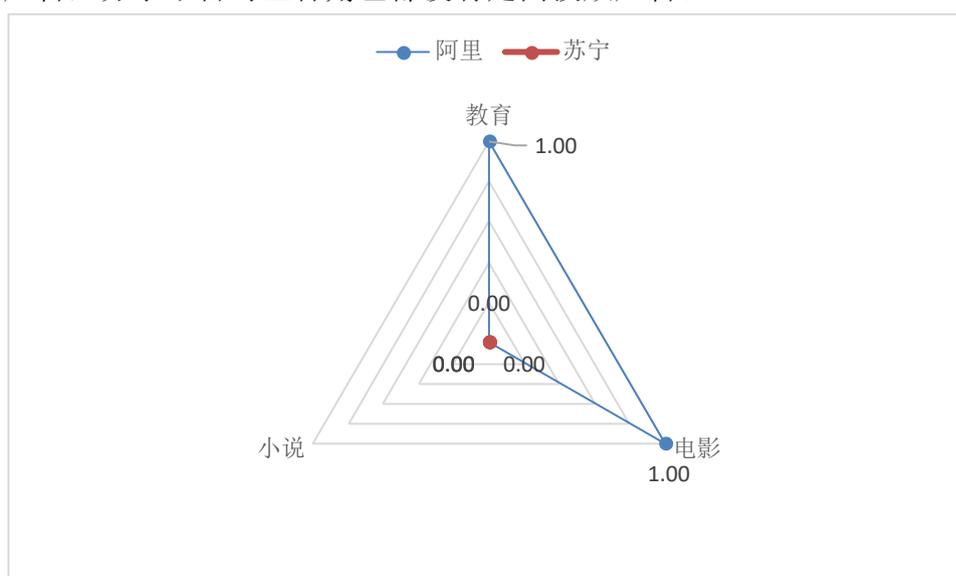

图 4-11 阿里与苏宁 TTK 值

Figure 4-11 TTK of Ali and Suning

也就是说，同样是为自身电商平台商品做推广，阿里 Tanx 平台能够针对用户偏好投放具有关键词或者与该偏好相关的定向广告，而苏宁广告平台没有这种能力。

### 4.2.4 不同 ADX 平台的时间动态行为的对比测量及分析

由上一小节可知百度、谷歌和阿里广告平台都可以针对用户偏好定向投放广告，接下来本文希望研究这几个平台判断出用户偏好所需要的时间以及能够记住

该用户偏好的时间长短。

### 4.2.4.1　两个平台广告投放的时间动态行为对比

　　BAiLP 指标捕获了广告落地页关键词集合与训练网页关键词的交集。当空白角色收到的广告包含训练网页关键词时，需要先排除这部分广告的影响。换句话说，BAiLP 代表了由虚拟角色兴趣 i 带来的行为定向。

　　图 4-12 和图 4-13 是谷歌和百度十天内的 BAiLP 值，它们分别反映了谷歌和百度平台针对"金融"和"设计"两种虚拟角色广告投放的时间动态行为。红色箭头表示从这一天起改变虚拟角色的兴趣特征。

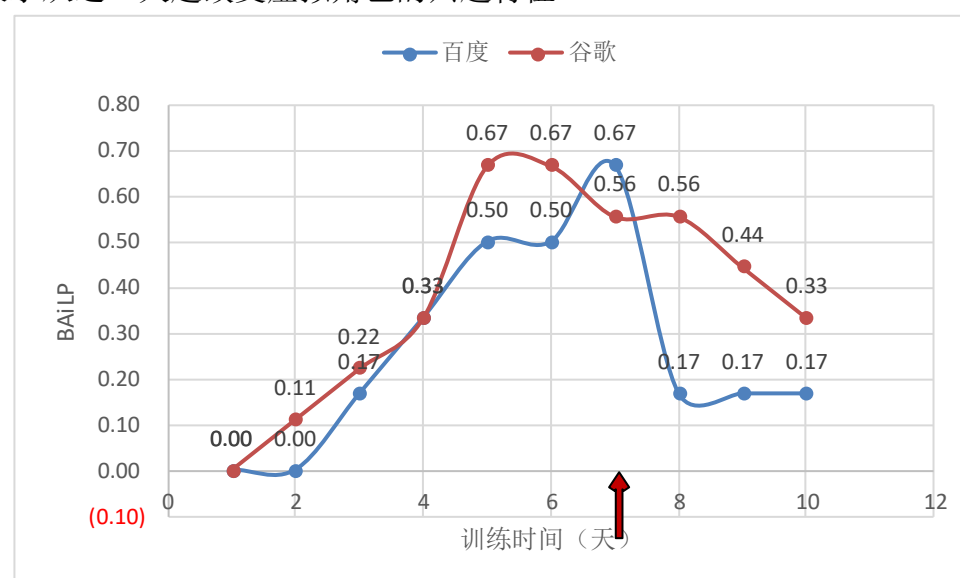

图 4-12 百度与谷歌金融角色 BAiLP 值

Figure 4-12 BAiLP value of "financial" persona on Baidu and Google platform

　　由图 4-12，可以观测到谷歌平台从测量第 2 天开始就针对"金融"角色定向投放广告，且增速明显，直到第 5 天达到最大值 0.67，定向效果明显；而百度平台则是在训练第 3 天才对出现针对该角色的定向广告，由此可知道谷歌平台对角色兴趣特征的识别比较迅速。在第 7 天将"金融"角色改变为"教育"角色，百度平台针对"金融"角色的定向广告迅速减少，而谷歌平台衰减速度较慢，由此可知谷歌平台对虚拟角色的兴趣特征记忆比较久，遗忘速度慢。

　　由图 4-13，谷歌平台第 2 天开始对"设计"角色定向投放广告，到第 5 天达到最大值 0.13，有一定定向效果；百度平台从第 3 天才开始定向投放广告，但是定向效果比较显著，到第 7 天达到了 0.40。在第 5 天将"设计"角色改为"旅游"角色，谷歌平台从第 6 天开始就没有针对"设计"角色的定向广告了；而百度平台针对"设计"角色的定向广告还在增加。

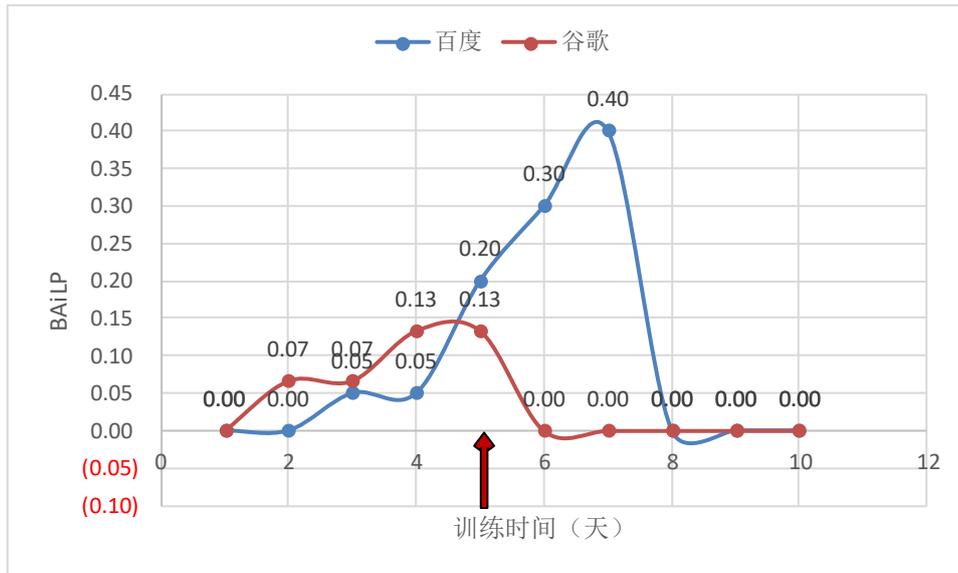

图 4-13 百度与谷歌设计角色 BAiLP 值

Figure 4-13 Baidu and Google design role BAiLP value

综上所述，通过对百度和谷歌两个平台广告投放的时间动态行为进行对比分析，能够发现谷歌对虚拟角色的兴趣特征把握比较敏锐，对于"金融"这类行为定向效果显著的角色记忆比较久，对于"设计"这类行为定向效果不明显的角色遗忘速度快；百度平台对虚拟角色的兴趣特征反应不及谷歌平台灵敏，但其定向效果好，且对角色的记忆比较久。

### 4.2.4.2　单平台广告投放时间动态行为

（1）灵敏程度

根据 3.3 节对虚拟角色构建的方法介绍，将带有某一偏好标签的训练网页和收集广告的门户网站放在一个列表里，通过随机访问列表里的网页同时进行虚拟角色训练和广告收集两项任务。根据这个过程中收集到的广告，可以分析各个广告平台对用户偏好做出判断的敏锐性，也就是它们需要多少时间才会发现该用户的偏好。

图 4-14 和图 4-15 分别是训练金融和设计虚拟角色过程中百度广告平台的广告投放情况，这里的第一天表示的是开始训练虚拟角色的时间，红色箭头标记的是指会在这天改变虚拟角色的兴趣特征。从图 4-14 中可以看出从第四天起百度平台金融类别的广告有明显增长，直到第七天的时候达到最大占比 10%。图 4-15 则显示第三天的时候百度平台艺术类广告开始变多，并在接下来几天逐渐增长。

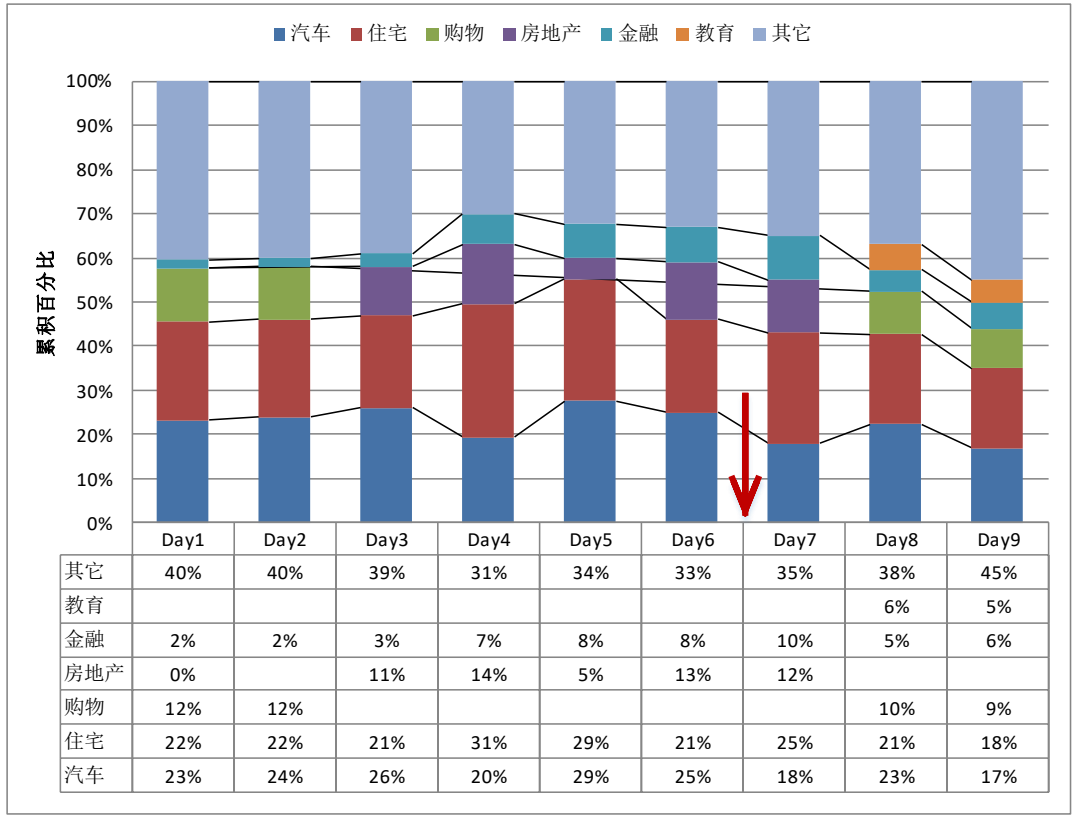

图 4-14 "金融"角色收到的百度广告

Figure 4-14 Advertising received by "finance" role from Baidu ADX

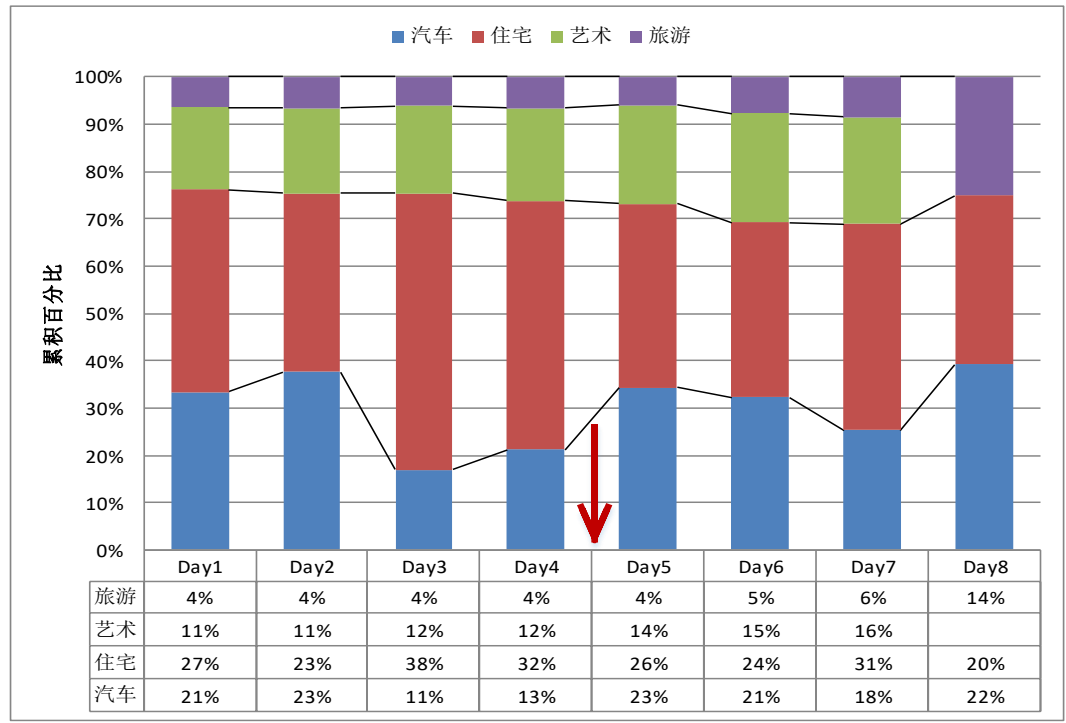

图 4-15 "设计"角色收到的百度广告

Figure 4-15 Advertising received by "design" role from Baidu ADX

图 4-16 和图 4-17 分别是训练金融和设计虚拟角色过程中谷歌广告平台的广告

投放情况。从图 4-16 可以发现谷歌平台一开始就有 30%的金融类别广告，且从第二天起该类别广告一直呈增长趋势。而图 4-17 反映出谷歌广告平台在第二天就投放了艺术类别的广告。

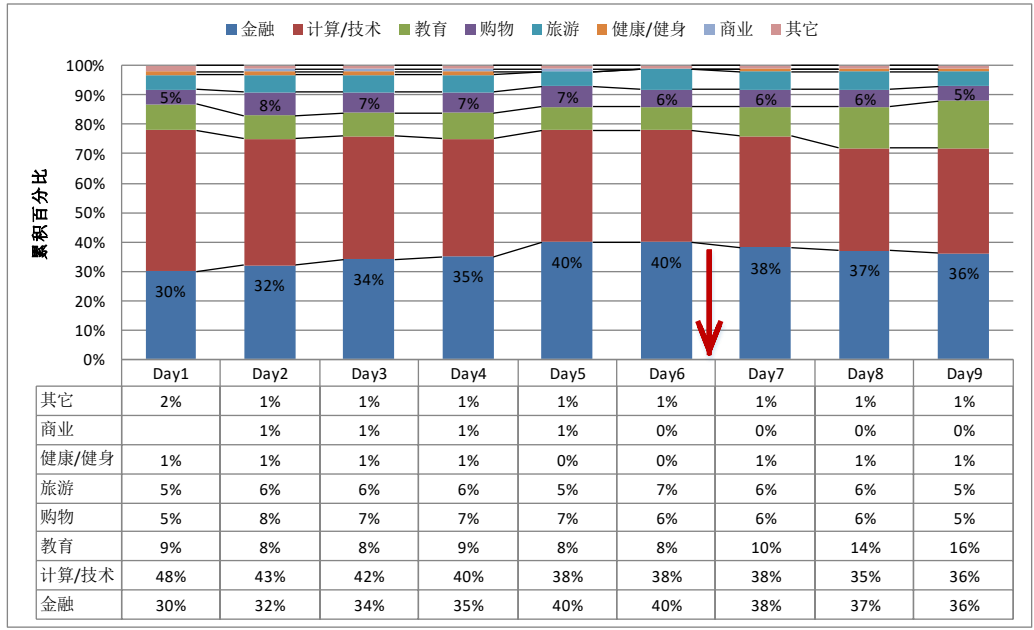

图 4-16 "金融"角色收到的谷歌广告

Figure 4-16 Advertising received by "finance" role from google ADX

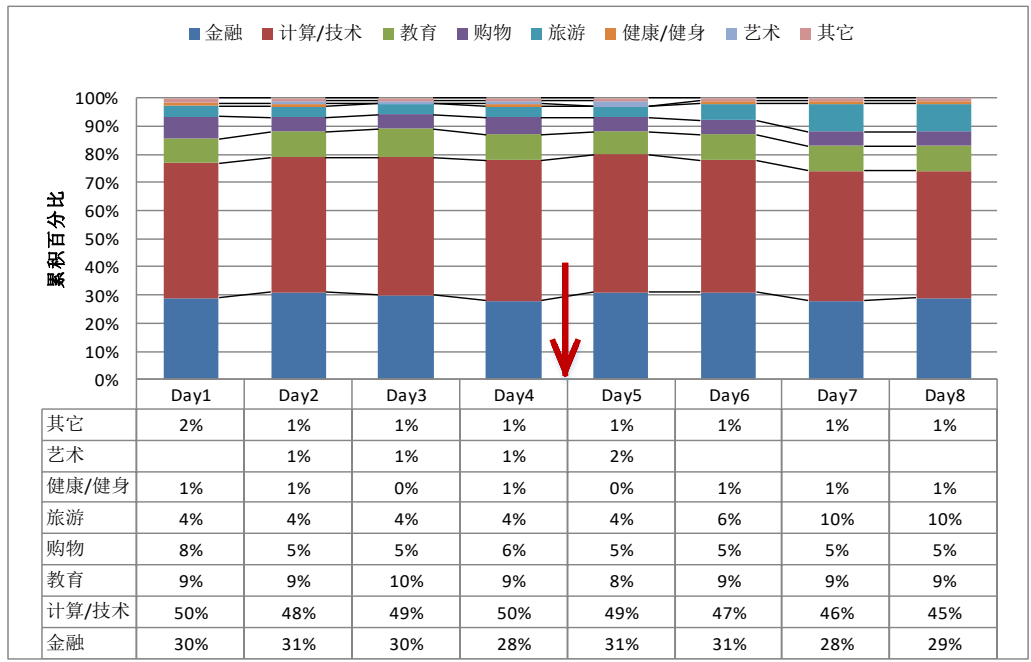

图 4-17 "设计"角色收到的谷歌广告

Figure 4-17 Advertising received by "design" role from google ADX

经过对比分析，百度广告平台判断用户偏好需要三至四天的时间，而谷歌平台则在第二天就能对用户偏好做出反应，说明谷歌广告平台对用户偏好的把握更加敏锐。

（2）记忆性

分析了广告平台定向广告效果后还有两个问题：当广告平台识别出用户偏好之后它会一直针对该偏好投放广告吗？如果用户偏好改变了广告平台能快速做出相应的改变吗？为了回答以上问题，在训练虚拟角色过程中将"金融"和"设计"类的训练网页替换为"教育"和"旅游"类别。要说明的是替换类别的选择是考虑到与之前类别相差较大，且对行为定向广告影响程度相似的。

在训练虚拟角色的第 7 天将百度平台训练网页从金融类变为教育类，图 4-14 中第 8 天的数据显示金融类别的广告骤减至 5%，而教育类的广告出现并达到 6% 的占比。而图 4-15 中第 8 至 10 天设计类别的广告已完全消失，旅游类别广告快速出现并占比百分之十几。由此可以看出，对于金融这一类对行为定向广告影响较大的偏好，百度广告平台不会快速遗忘但相应地广告投放会减少，而对于设计这类影响较小的偏好其遗忘速度很快且彻底。另一方面，针对用户行为偏好的变化百度广告平台能迅速做出反应并定向投放广告。

对于谷歌广告平台在第 5 天改变了虚拟角色类别，图 4-16 中第 6 天金融类别的广告并没有减少，之后几天才缓慢地递减。相反的，图 4-17 显示第 6 天设计类别的广告就快速消失了。同样的教育类和旅游类的广告也有较大的增长。因此可以判断出谷歌对金融这一类别的偏好记忆性较强，而对设计这一类别的偏好定向广告投放时间很短。

## 4.3　本章小结

本章主要利用第三章提出的多 ADX 平台测量系统对百度、谷歌等多个 ADX 平台进行性能评估，并发现几点结论。首先，各个广告平台对投放广告的网页有不同倾向。百度、谷歌以及搜狗三个广告交易平台最青睐的都是计算/技术类网页，且投放比例都达到 20%以上。其次，通过对百度、谷歌以及阿里、苏宁进行两两对比，发现相比于谷歌交易平台，百度交易平台的定向广告投放能力更强，具体体现在它除了能针对用户现有偏好投放广告还可以智能联想用户的其它兴趣领域。同样是为自身电商平台商品做推广，阿里 Tanx 平台能够针对用户偏好投放具有关键词或者与该偏好相关的定向广告，而苏宁交易平台没有这种能力。最后，还发现谷歌交易平台对用户偏好的把握更加敏锐，对用户偏好的记忆性和百度交易平台类似。

# 5 结论

随着互联网的普及与快速发展，它对人们日常生活有着越来越深入的影响，在线广告已经成为目前主流广告投放形式。

广告交易平台作为连接广告主与媒体主的桥梁，其投放广告的性能对广告主和媒体主都至关重要。广告交易平台的主要职责是将广告主的商品广告投放到媒体主的广告位上，既能帮助广告主推广商品又可以将媒体主网络流量变现。为了增加广告转化率并提升广告主收益，广告交易平台通常会根据访问媒体网站的用户行为偏好定向投放广告。广告交易平台定向投放广告的能力不仅影响广告主的收益，也会给用户带来不一样的上网体验，从而影响网络媒体的吸引度。

本文首先介绍了在线广告系统架构，阐述了该系统的各参与者及相互关系。在此基础上，本文整理了在线广告系统的相关研究并指出其局限性。

本文提出了同步跨平台测量方法，在该方法中，提出一个多 ADX 平台性能比较的测量基准；提出了并行访问广告链接方法。为了实现该方法，本文设计开发了一个同步跨平台测量系统。

为了评估本文设计开发的同步跨平台测量系统的有效性，利用该系统对实际的 ADX 平台进行了测量和比较分析。实验结果达到了预期效果，在同一基准下，该系统能够明显区分出不同 ADX 平台对相同虚拟角色的广告投放差异；并发现不同ADX平台广告投放会随虚拟角色改变而变化，且这种变化的灵敏程度存在不同。

基于本文研究，后续将设计大规模广告网络测量系统，并基于测量数据对系统进行理论建模分析。

# 参考文献